\documentclass[pdflatex,sn-mathphys]{sn-jnl}%

\usepackage{algorithm}
\usepackage{graphicx}
\usepackage[font=footnotesize,labelfont=bf]{caption} 
\usepackage{subcaption}
\usepackage{array}
\usepackage{multirow}
\theoremstyle{thmstyleone}%
\theoremstyle{thmstyletwo}%

\theoremstyle{thmstylethree}%

\newcommand\SD{Slater determinants }
\newcommand{\ket}[1]{ \vert #1 \rangle }
\newcommand{\bra}[1]{ \langle #1 \vert }

\newcommand{\twodimmat}[4]{
	\left[\!
	\begin{array}{cc}
		#1 & #2\\
		#3 & #4
	\end{array}
	\!\right]}

\newcommand{\ceil}[1]{\lceil #1 \rceil}
\newcommand{\ssc}{\vec{s}}
\newcommand{\qesci}{VQ-SCI }

\newcommand{\vqfci}{VQ-FCI }

\usepackage{xcolor}     
\usepackage[normalem]{ulem}
\usepackage{color,soul}

\begin{document}

\title[A Qubit-Efficient Variational Selected Configuration-Interaction Method]{A Qubit-Efficient Variational Selected Configuration-Interaction Method}

\author[1,2]{\fnm{Daniel} \sur{Yoffe}}\email{d4444i@gmail.com}

\author[1]{\fnm{Amir} \sur{Natan}}\email{amirnatan@tauex.tau.ac.il}

\author[2,3]{\fnm{Adi} \sur{Makmal}}\email{adi.makmal@biu.ac.il}

\affil[1]{\orgdiv{Department of Physical Electronics}, \orgname{Tel Aviv University}, \orgaddress{\city{69978 Tel Aviv}, \country{Israel}}}

\affil[2]{\orgdiv{The Engineering Faculty}, \orgname{Bar-Ilan University}, \orgaddress{\city{52900 Ramat-Gan}, \country{Israel}}}

\affil[3]{\orgdiv{The center for Quantum Entanglement Science and Technology}, \orgname{Bar-Ilan University}, \orgaddress{\city{52900 Ramat-Gan}, \country{Israel}}}

\abstract{
Finding the ground-state energy of molecules is an important and challenging computational problem for which quantum computing can potentially find efficient solutions. 
The variational quantum eigensolver (VQE) is a quantum algorithm that tackles the molecular groundstate problem and is regarded as  
one of the flagships of quantum computing. Yet, to date, 
only very small molecules were computed via VQE, due to high noise levels in current quantum devices. 
Here we present an alternative variational quantum scheme that requires significantly less qubits. 
The reduction in qubit number allows for shallower circuits to be sufficient, rendering the method more resistant to noise.
The proposed algorithm, termed  variational quantum selected-configuration-interaction (VQ-SCI), is based on:
(a) representing the target groundstate as a superposition of Slater determinant configurations, encoded directly upon the quantum computational basis states; and (b) selecting \emph{a-priory} only the most dominant configurations.  
This is demonstrated through a set of groundstate calculations of the H$_2$, LiH, BeH$_2$, H$_2$O, NH$_3$ and C$_2$H$_4$ molecules in the sto-3g basis set, performed on IBM quantum devices. We show that the VQ-SCI reaches the full-CI (FCI) energy within chemical accuracy using the lowest number of qubits reported to date.  
Moreover, when the SCI matrix is generated ``on the fly", the VQ-SCI requires exponentially less memory than classical SCI methods. This offers a potential remedy to a severe memory bottleneck problem in classical SCI calculations. Finally, the proposed scheme is general and can be straightforwardly applied for finding the groundstate of any Hermitian matrix, outside the chemical context.}

\keywords{variational quantum algorithms, variational quantum eigensolver, qubit-efficient, electronic structure, configuration interaction, first quantization}

\maketitle

\section{Introduction}\label{sec1}
    Finding the electronic structure of a chemical system is at the heart of computational chemistry with countless applications in material design, drug design, and scientific research \cite{dykstra2011theory, szabo2012modern, hoffman2014mechanism, aspuru2018matter,martin2020electronic}. 
     This requires the solution of the non-relativstic time-independent Schr\"{o}dinger equation:
    \begin{equation}
    \label{eq:shroedinger}
        \mathcal{H}\ket{\Psi} = \mathcal{E}\ket{\Psi}, 
    \end{equation}
    where $\mathcal{H}$ is the system's Hamiltonian, $\mathcal{E}$ is the total energy, and $\Psi$ is the corresponding groundstate, dictating the electronic structure of the system.  
    However, solving the electronic Schr\"{o}dinger equation directly is difficult because of the enormous dimensionality of the problem, which grows rapidly with the number of electrons. As relatively small molecules already encapsulate tens of electrons, e.g., water ($N\!\!=\!\!10$) or benzene ($N\!\!=\!\!42$), finding an exact solution becomes intractable, even for very strong classical computers \cite{saad2010numerical}.

    The variational quantum eigensolver (VQE) is a recent quantum algorithm that aims at solving the electronic structure problem in polynomial timescales \cite{peruzzo2014variational_org_vqe,mcclean2016theory, mcardle2020quantum_comp_chemistry}. 
    It is based on a parameterized quantum circuit that is 
    tuned iteratively until a minimum energy value is attained. The central advantage of VQE, in comparison to previous quantum algorithms (e.g.\ phase-estimation \cite{abrams1997simulation, abrams1999quantum}), is that it requires shallower circuits which are more noise tolerant. It is therefore better suited for the current noisy intermediate scale quantum (NISQ) devices and is considered to be a prime candidate for demonstrating quantum advantage. 
    
    Yet, despite its promise, VQE has reached only limited success so far.
    Real quantum hardware VQE demonstrations are still scarce, addressing only small molecules, and reaching insufficient accuracy even for the minimal basis set, see \cite{peruzzo2014variational_org_vqe,o2016scalable,shen2017quantum, kandala2017hardware,google2020hartree, nam2020ground_watre_vqe_trapped_ion, yeter2021benchmarking, barison2022quantum} for central examples.  
    Moreover, it was estimated that for practical usage VQE would require millions of two-qubit gates and hundreds of \emph{logical}, error-free, qubits \cite{kuhn2019accuracy}, much beyond current capabilities. 
    The reason is that VQE in the standard Jordan-Wigner (JW) mapping \cite{jordan1993paulische}, requires a number of qubits that equals the number of spin-orbitals ($M$) in a given basis set (assuming no core orbital freezing and symmetry reductions, see below for a formal description), which grows with the system. As the number of qubits increases, more gates are required for the variational circuits, which despite being relatively shallow, still accumulate too much noise. Furthermore, even on noiseless simulations, VQE could so far address only small molecules \cite{kuhn2019accuracy, nam2020ground_watre_vqe_trapped_ion, cao2021towards} due to its large qubits consumption, which is difficult to simulate.

    Several attempts have been made to improve the VQE through computational resource saving and better noise mitigation \cite{mcclean2016theory, tilly2021variational_vqe_best_practices}. For example, 
    some suggestions have been made to reduce circuit's depth by replacing the original chemically inspired Ansatz with more heuristic approaches, see e.g.\ \cite{kandala2017hardware,grimsley2019adaptive_vqe,benfenati2021improved,ostaszewski2021reinforcement}, which are often prone to particle number and spin symmetry violation. In addition, several attempts to employ better basis sets have been made \cite{barison2022quantum, hong2022accurate} which led to a more accurate description of the Hamiltonian, thus reaching better accuracy at a comparable computational cost. 
    Reducing the number of distinct Pauli strings required to be  measured was also pursued in \cite{kandala2017hardware, verteletskyi2020measurement,huggins2021efficient}.

    Resource saving via qubit reduction was also explored, for example through core orbital freezing (COF)~\cite{eddins2022doubling, barison2022quantum}, which, however, introduces an uncontrolled approximation. 
    An alternative direction towards qubit reduction, which is also the one we explored, employs a different information encoding, as described next.

    \paragraph{Information encoding}
    The computational resources required for calculating  molecular systems are usually measured in terms of the number of electrons in the system, $N$, and the size of the basis-set, namely the number of spin-orbitals that are taken into account, $M$. 
    VQE is formulated within the framework of second quantization \cite{tilly2021variational_vqe_best_practices}, where each Slater determinant is encoded as a binary string of spin-orbital occupations (see Sec.~\ref{sec:background}). Accordingly, in the standard JW encoding \cite{jordan1993paulische} the quantum state of each qubit encodes the occupation of the corresponding spin-orbital, therefore requiring $M$ qubits, one per spin-orbital. 
    Several studies aimed at reducing this qubit requirement by relying on symmetry considerations, e.g.\ $Z_2$-symmetry based qubit tapering \cite{bravyi2017tapering} and qubit reduction through molecular point group symmetry \cite{setia2020reducing}. Another attempt was explored in the form of ``entanglement forging" \cite{eddins2022doubling}, where a system of $2N$ qubits is regarded as two, weakly entangled, $N$ qubit systems, thus effectively cutting the number of qubits by half, at a cost of additional classical calculations. Alternatively, when working in the framework of first quantization, other types of encoding may be useful. 
    For example, in the early work of Abrams and Lloyd~\cite{abrams1997simulation} it was suggested to simulate a chemical system using $N\log{M}$ qubits, by 
    encoding the spin-orbital of each electron as a separate binary string of length $\log{M}$. This encoding was later explored in \cite{berry2018improved}, see also \cite{mcardle2020quantum_comp_chemistry}.

    A different encoding associates the $k$'th Slater determinant
    with the computational state $\ket{bin(k)}$, where $bin(k)$ is a predefined decimal-to-binary mapping of $k$. Such an encoding, denoted \emph{amplitude encoding} in the quantum machine learning literature where it is often used, see \cite{schuld2018supervised}, or simply \emph{binary encoding},  fully utilizes the capacity of $q$ qubits to encode $2^q$ amplitudes and is therefore economical in the number of required qubits. Examples for using it in the chemical context include quantum simulation \cite{toloui2013quantum, babbush2017exponentially, bravyi2017tapering}, simulating molecular vibrations of a bosonic operator \cite{mcardle2019digital_qubit_efficient_molecular_vibrations}, $d$-level quantum operator (qudits) encodings \cite{sawaya2020resource_qubit_efficient_d_level}, and finding deuteron ground-state energy \cite{di2021improving_qubit_efficient_Grey_code}.

    A recent paper suggested using such a binary encoding for finding the molecular groundstate \cite{QEE_shee2022qubit} within the variational quantum paradigm. 
    Essentially, this amounts to solving the molecular groundstate problem in the representation of Slater determinants rather than that of spin-orbital occupations. 
    It enables a FCI calculation with $\log D_{FCI}$ qubits, where $D_{FCI}=\binom{M}{N}$ is the number of \SD that can be composed within a given basis-set.
    Note that $D_{FCI}=\binom{M/2}{N/2}^2$ when the calculation is spin restricted to equal number of spin up and down electrons $N_{\uparrow}=N_{\downarrow}$ (but without imposing the same spatial orbital for spin up and spin down). In this paper, we use such a spin-restriction assumption, for all the presented analyses. 
    
    For a fixed number of electrons and increasing number of spin-orbital candidates, this encoding requires a number of qubits that scales only logarithmically with the number of spin-orbitals, providing an exponential qubit saving over the standard VQE formalism. Indeed, it was demonstrated in \cite{QEE_shee2022qubit} that $H_2$ could be solved to high accuracy with the 6-31G basis set using merely 4 qubits rather than 8, when solved in the basis of Slater determinants. An additional, more qualitative advantage of this scheme is that it leads, by construction, to groundstates that preserve spin and electron number, irrespective of the quantum circuit's Ansatz, thus allowing the usage of hardware-efficient Ansatz with no computational artifacts, as described in \cite{QEE_shee2022qubit}.

    The problem, however, with performing such a variational quantum FCI calculation directly in the basis of \SD is that it requires using all slater determinants to encode the Hamiltonian and therefore the measurement of $\mathcal{O}(D_{FCI}^2)$ distinct Pauli strings (see \cite{QEE_shee2022qubit} and Sec.~\ref{sec:resource_analysis}). This exerts strict limitations on the practical usage of such an approach to (yet again) small molecules. Moreover, the saving in qubit number is mild. For example, the ethylene molecule (C$_2$H$_4$, $N=16$ electrons) has $M=28$ spin-orbitals in the minimal sto-3g basis set, hence requiring $\ceil{\log_2 D_{FCI}}=\ceil{\log_2{\binom{14}{8}\binom{14}{8}}}=24$ qubits, even under spin-restriction,  compared to $M=28$ qubits in the standard VQE. 
    For such a calculation to be feasible the number of \SD that take part in the calculation must therefore be significantly reduced.

    \paragraph{Full vs. selected-CI}
    One way to reduce the number of \SD is to perform a selected-CI (SCI) computation \cite{tubman2020modern}. 
    In this case, the computation is performed on a selected, smaller set, of Slater determinants, instead of the full one, thereby requiring much smaller, yet sparse, matrices to be diagonalized, significantly reducing the computational cost. The SCI scheme has been studied extensively as a classical computational method for half a century, see e.g., \cite{bender1969studies_org_SCI, huron1973iterative, buenker1978applicability_MRDCI, evangelisti1983convergence_improved_CIPSI, harrison1991approximating_SCI_PT} for some of the first studies that developed heuristics for selecting the best $k$  \SD candidates. 
    They showed that chemical accuracy can be reached with a relatively small subset of Slater determinants, when selected wisely.   
    In recent years, substantial progress has been made with a wide variety of new methods for selecting the most important Slater determinants, using advanced tools from various approaches, e.g.,
    localized orbitals \cite{ben2011direct_SCI_licalized_orbitals}, perturbation theory \cite{giner2013using_SCI_PT}, energy cut-off \cite{evangelista2014adaptive_SCI_energy_cutoff}, heat-bath sampling approach \cite{holmes2016heat_SCI_heatbath_sampling}, approximating the amplitudes of the ground-state wavefunction \cite{tubman2016deterministic_SCI_choosing_SD_deterministically}, iterative selection \cite{zhang2020iterative}, and machine learning \cite{pineda2021chembot_SCI_ML, goings2021reinforcement_SCI_RL}. These developments led to more scalable and accurate results, 
    making the classical SCI method a competitive approach for conducting electronic structure calculations with low computational costs.
    A recent study  \cite{feniou2023overlap} even considered the usage of a classically computed SCI wavefunction as a target wavefunction for better preparing an initial Ansatz for the ADAPT-VQE scheme \cite{grimsley2019adaptive_vqe}.

    Performing a SCI calculation by means of quantum computing was not yet sufficiently explored. Moreover, executing SCI within the VQE approach, namely constraining the VQE calculation to \emph{any} desirable subset of Slater determinants, is not straightforward. It is 
    possible to constraint the VQE to a specific set of excitations, 
    using a chemically inspired Ansatz,  see e.g.~\cite{nam2020ground_watre_vqe_trapped_ion}, 
    but this requires a deep circuit, which accumulates much noise.

\begin{figure}[h!]
    \centering
    \includegraphics[width=0.5\textwidth]{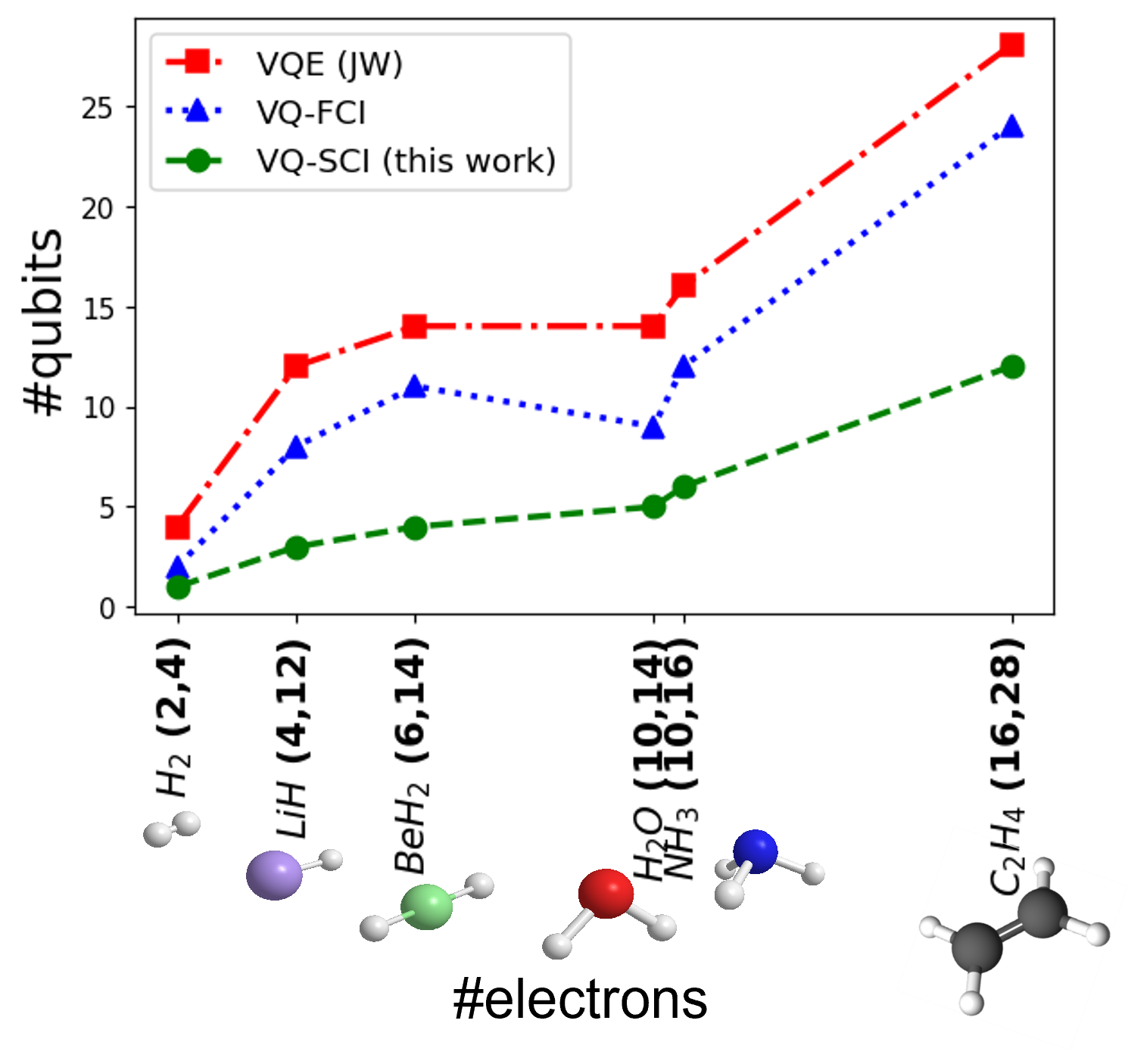}
    \caption{
    The number of qubits required to perform groundstate energy calculations of the molecules studied in this paper, ordered by their number of electrons (the NH$_3$ molecule is slightly shifted for clarity), in the sto-3g basis set, per method. The number of electrons and spin-orbitals $(N,M)$ are written next to each molecule. VQE assumes JW encoding (red squares), VQ-FCI employs all possible Slater determinants (blue triangles), and the proposed VQ-SCI (green circles) employs a small subset of the most significant \SD that lead to the FCI accuracy level, see details in text. The VQ-SCI is shown to be the most economical in the number of required qubits.} 
    \label{fig:qubits_per_electrons}
\end{figure}
    
\subsection{Our approach}
\label{sec:our_approach}
    Here we develop and explore the idea of solving the electronic structure problem with the aid of quantum computers, by performing variational quantum selected-configuration-interaction (VQ-SCI) calculation, directly in the basis of Slater determinants.
 
    Building upon classical SCI schemes, we perform the selection of $k$ significant \SD classically, but use quantum computers to find the groundstate of the corresponding SCI matrix. 
    By performing the calculation in the \SD basis and encoding the \SD to qubits via a decimal-to-binary encoding, our VQ-SCI method is very modest in qubit consumption.    
    This allows us to perform groundstate energy calculations for a set of small to medium-size molecules, within the sto-3g basis set, with the smallest number of qubits reported to date, while addressing all electrons in the system and without resorting to any kind of orbital freezing. In particular, we calculated the groundstate of the following molecules: H$_2$ ($N=2, M=4$, $q=1$ qubit), LiH ($N=4, 
 M=12$, $q=3$ qubits), BeH$_2$ ($N=6, M=14$, $q=4$ qubits), H$_2$O ($N=10, M=14$, $q=5$ qubits), NH$_3$ ($N=10, M=16$, $q=6$ qubits), and C$_2$H$_4$ ($N=16, M=28$, $q=12$ qubits), 
 where $q$ is the number of qubits required by \qesci to reach the FCI groundstate energy within chemical accuracy.

    To demonstrate the saving in qubits number, Fig.~\ref{fig:qubits_per_electrons} presents the number of qubits required by standard VQE in the JW encoding (red squares) and by our VQ-SCI approach (green circles). For completeness, the number of qubits required by a VQ-FCI method (blue triangles), i.e.\ when using \emph{all} \SD (similar to the calculations presented in \cite{QEE_shee2022qubit}), is also depicted. The figure presents all the molecules we addressed in this paper, ordered by their number of electrons. In all cases the sto-3g basis set is considered and no core orbital freezing is assumed. For the VQE scheme in the JW encoding, $M$ qubits are needed for $M$ spin-orbitals. The VQ-FCI requires $log (D_{FCI})$, where $D_{FCI}=\binom{M/2}{N/2}^2$ is the number of possible Slater determinants. Finally, our proposed VQ-SCI requires $log (D_{SCI})$ qubits, where $D_{SCI}$ is the number of significant Slater determinants required to reach the FCI groundstate energy within chemical  accuracy. The exact value of $D_{SCI}$ depends on the chosen selection procedure.  The one we employed is described in Sec.~\ref{sec:results} and reaches $D_{SCI} \approx \sqrt{D_{FCI}}$, thus cutting the number of qubits approximately by half (often more). For both VQ-FCI and VQ-SCI we assumed spin-restriction, as described above. 
    It is seen that while VQ-FCI improves the number of qubits only mildly in comparison to standard VQE, the VQ-SCI requires much fewer qubits and increases much slower than the VQE. Evidently, it requires a number of qubits that is always below the number of electrons in the molecule.

    Consequent advantages of the \qesci method in comparison to VQE are: (a) The use of fewer qubits makes shallower circuits sufficient, which  translates to a significant reduction in the circuit's noise. This, in turn, leads to better accuracy; (b) The groundstate Ansatz is constructed as a linear combination of only physical states, which leads to the conservation of total particle number and spin-symmetry, by construction; (c) Selecting the \SD does not come at the cost of using deep chemical inspired circuit Ans\"atze; and (d) Although the method selects a specific set of Slater determinants, it does not perform any sort of orbital ``freezing". This ability to choose the \SD without being restricted to a specific set of orbitals gives the VQ-SCI method more degrees of freedom, which allow it to include excitations that are often dismissed. These advantages come at the cost of more Pauli strings needed to be measured. We analyze this cost, point to cases where the number of Pauli strings in VQE is nevertheless larger than in the VQ-SCI method, and discuss different directions to reduce it. 

    Most importantly, the proposed method constitutes a general procedure for finding the lowest eigenvalue and groundstate of a $D$ by $D$ Hermitian matrix, using $\log D$ qubits. In fact, we are only constrained to matrices whose lowest eigenvalue is real (but are not necessarily Hermitian). 
    The scheme can therefore be applied outside the chemical context, for example, in graph theory, where the smallest eigenvalue of the adjacency matrix entails important information on the graph's structure \cite{cioabua2019some_smallest_eigenvalue_graph_theory}. Our analysis indicates that in comparison to common classical approaches, the quantum scheme scales similarly with time and logarithmic in memory. This advantage can be achieved only when the calculation is performed on actual quantum hardware and can not be reproduced via classical simulations.

The paper is structured as follows: Sec.~\ref{sec:background} describes the necessary background of computational chemistry and the variational quantum eigensolver (VQE) algorithm;
Sec.~\ref{sec:methods} explains the formalism of the proposed VQ-SCI approach, including a detailed example for the case of H$_2$; Sec.~\ref{sec:results} presents our computational VQ-SCI results, displaying groundstate energy calculations for H$_2$, LiH, BeH$_2$, H$_2$O, NH$_3$, and C$_2$H$_4$ in the sto-3g basis set, through noiseless, noisy, and real-hardware calculations. Sec.~\ref{sec:resource_analysis} offers a crude resource analysis of the \qesci method and discusses its potential applicability. Sec.~\ref{sec:conclusion} concludes the paper and points to future research directions.

\section{Background}\label{sec:background}
\subsection{Classical computational chemistry}
    The Hamiltonian of a molecular system of $A$ atoms with $N$ electrons takes the following form (under the Born-Oppenheimer approximation) \cite{dykstra2011theory, szabo2012modern}: 
    \begin{equation}
        \label{eq:molecular_Hamiltonian}
        \mathcal{H} = \sum_{j=1}^N T_j+\sum_{a=1}^A \sum_{j=1}^N V_{a,j}^{ext}+\sum_{\substack{j=1, \\ k>j}}^N V_{j,k}^{e-e}, 
    \end{equation}
    where $T_j = -\frac{\nabla_j^2}{2}$ is the kinetic energy of a single electron, $V_{a,j}^{ext} = -\frac{Z_a}{\vert \vec{R}_a - \vec{r}_j \vert}$ is the external electronic nuclei-electron attraction, and $V_{j,k}^{e-e} = \frac{1}{\vert \vec{r}_j -\vec{r}_k \vert}$ accounts for the electron-electron interactions.  Atomic units are used throughout. 
    
    The Hartree-Fock (HF) method provides an approximated solution for finding the groundstate of the Schr\"odinger Eq.~(\ref{eq:shroedinger}) with the molecular Hamiltonian of Eq.~(\ref{eq:molecular_Hamiltonian}) in the form of a single Slater determinant wavefunction:
    \begin{equation}
        \label{eq:SD}
        \Phi(\ssc_1,\hdots,\ssc_N) = \frac{1}{\sqrt{N!}}
    \begin{vmatrix}
    \chi_1(\ssc_1) & \chi_2(\ssc_1) & \hdots & \chi_N(\ssc_1)\\
    \chi_1(\ssc_2) & \chi_2(\ssc_2) & \hdots & \chi_N(\ssc_2)\\
    \vdots & \vdots & \vdots\\
    \chi_1(\ssc_N) & \chi_2(\ssc_N) & \hdots & \chi_N(\ssc_N)
    \end{vmatrix} \equiv SD\{\chi_1,...,\chi_N\} 
    \quad, 
    \end{equation}
    where $\{\chi_j(\ssc_k)\}$ are a set of molecular orbitals, $SD\{\chi_1,...,\chi_N\}$ mark the Slater determinant composed of the $\{\chi_1,...,\chi_N\}$ molecular orbitals, and $\ssc$ stands for the spatial $(\vec{r})$ and spin $(\alpha)$ coordinates of a single electron. 
    The HF approximation is the simplest, hence most efficient, approximation to an uncorrelated $N$-electron wavefunction, which satisfies the antisymmetry requirement. 
    Yet, it performs rather poorly, in terms of the achievable groundstate energy. 
   
    \paragraph{Configuration interaction (CI)}
    One way to go beyond the HF approximation is via the configuration-interaction (CI) method, which represents the molecular groundstate as a linear combination of many Slater determinants (rather than just a single one): 
    \begin{equation}
        \label{eq:CI_state}
        \ket{\Psi(\ssc_1,\hdots,\ssc_N)} = \sum_{k}{c_k \ket{\Phi_k(\ssc_1,\hdots,\ssc_N)}}.  
    \end{equation} 
    It is guaranteed by the variational theorem that taking a larger-size set of \SD (or, synonymously, configurations) necessarily brings us closer to the desired groundstate wavefunction. However, reaching absolute accuracy might well require an intractable large set of Slater determinants. 
    A common way to bound the calculation to a manageable number of \SD is to employ confined basis sets (such as the sto-3g) with finite sets of $M>N$ molecular spin-orbitals, allowing the composition of $F=\binom{M}{N}$ different Slater determinants \cite{hehre1969self_sto3g}. 
    When the summation in Eq.~\ref{eq:CI_state} accounts for all possible $F$ Slater determinants, we refer to the calculation as a Full-CI (FCI) calculation, whereas when a smaller subset of $D < F$ \SD is employed we refer to the calculation as a Selected-CI (SCI) calculation. 
    In either case, finding the groundstate of a molecule  amounts to finding the lowest eigenstate, $\Psi_0$, of the following Hermitian matrix: 
    \begin{equation}
        \label{eq:CI_matrix}
         M_{CI}\ket{\Psi_0} = e_0\ket{\Psi_0}, \quad M_{CI}(k,j) = \bra{\Phi_k}\mathcal{H}\ket{\Phi_j}, 
    \end{equation}
    which represents the molecular Hamiltonian operator $\mathcal{H}$ of Eq.~\ref{eq:molecular_Hamiltonian} in the basis of Slater determinants. 
    Reducing the number of \SD used in the calculation through SCI leads directly to reducing the dimension of the $M_{CI}$ Hamiltonian matrix. 
  
    The classical SCI method is a 2-phase algorithm: (I) Select $D$ \SD and construct the $D$-dimensional SCI matrix in the form of Eq.~\ref{eq:CI_matrix}; (II) Find the lowest eigenstate and eigenvalue of the SCI matrix. The second part of the algorithm is the computational bottleneck. It has a naive time complexity of $\mathcal{O}(D^3)$, which can be reduced to $\mathcal{O}(D^2)$ and an $\mathcal{O}(D)$ space complexity using iterative methods, such as the Davidson algorithm \cite{DAVIDSON197587}. 
    The proposed \qesci method, described in Sec.~\ref{sec:methods}, performs the second phase of the SCI algorithm by means of quantum computing.

    \subsection{Variational quantum eigensolver (VQE)}
    The VQE method \cite{peruzzo2014variational_org_vqe} searches for the system's groundstate by optimizing a parameterized quantum state $\psi(\theta)$ to  minimize the energy expectation value:
    \begin{equation}
    \label{eq:parameterized_energy}
        \Psi_0 = \arg\!\min_{\psi(\theta)} \bra{\psi(\theta)}\mathcal{H}\ket{\psi(\theta)},
    \end{equation}
    as dictated by the variational principle. 
    Here the molecular Hamiltonian takes the $2^{nd}$ quantization form:
    \begin{equation}
        \label{eq:second_quantization_Hamiltonian}
        \mathcal{H} = \sum_{p,q=1}^M h_{pq} a_p^\dagger a_q  
+   \sum_{p,q,r,s=1}^M\!\!\!h_{pqrs} a_p^\dagger a_q^\dagger a_r a_s,  
    \end{equation}
    where $M$ is the number of 
    spin-orbitals, and $a_p^\dagger$ and $a_p$ are the 
 fermionic creation and annihilation operators, respectively. 
    The scalar coefficients $h_{pq}$ and $h_{pqrs}$ are defined as follows, using the Hamiltonian parts of Eq.~\ref{eq:molecular_Hamiltonian}:
    \begin{equation}
    \begin{aligned}
        & h_{pq}=\bra{\chi_p}T_q+\sum_{a=1}^A V_{a,q}^{ext}\ket{\chi_q} = \int \chi^\ast_p(\vec{s}) \left(-\frac{\nabla_r^2}{2}  - \sum_{a=1}^A\frac{Z_a}{\vert \vec{R}_a - \vec{r} \vert} \right)  \chi_q (\vec{s}) d\vec{s} \\ 
        & h_{pqrs}=\bra{\chi_p\chi_q} V_{1,2}^{e-e} \ket{\chi_r\chi_s}=\int\int  \frac {\chi^\ast_p(\vec{s}_1)\chi^\ast_q(\vec{s}_2)\chi_r(\vec{s}_2)\chi_s(\vec{s}_1) }{\vert \vec{r}_1-\vec{r}_2 \vert} d\vec{s}_1 d\vec{s}_2,
        \end{aligned}
    \end{equation}
    where $\chi$ are molecular spin-orbitals of a single electron, as in Eq.~\ref{eq:SD}.

    The Hamiltonian of Eq.~\ref{eq:second_quantization_Hamiltonian} is encoded to the quantum circuit by mapping each fermionic operator to a string (tensor product) of single-qubit Pauli operators, resulting with the following summation: 
    $\mathcal{H} = \sum_k{a_k P_k}$, where $a_k$ are constant coefficients and $P_k = \bigotimes_{j=0}^{q-1}\sigma_{i,j}$ are strings of Pauli operators; the  $\sigma_{i,j}$ is the $i$'th Pauli operator ($i \in \{X,Y,Z,I\}$) applied onto the $j$'th qubit, and $q=M$ is the number of qubits in the circuit. Consequently, minimizing Eq.~\ref{eq:parameterized_energy} is performed by minimizing over $\sum_k{a_k}\bra{\psi(\theta)}P_k\ket{\psi(\theta)}$. The exact values of the $a_k$ coefficients as well as the specific choice of Pauli gates, depend on the employed fermionic to qubit operator mapping, e.g.\, Jordan-Wigner (JW) \cite{jordan1993paulische}, Bravi-Kitaev \cite{bravyi2002fermionic} and more (see \cite{mcardle2020quantum_comp_chemistry} and references therein), here we refer to the standard JW mapping.

    \section{The \qesci  Method}\label{sec:methods}    
    The proposed quantum variational SCI (VQ-SCI) algorithm is depicted in Alg.~\ref{alg:QESCI}. It deviates from classical SCI algorithms in that the second phase of finding the groundstate of the SCI matrix $M_{CI}$ is performed via quantum computing; 
    In what follows we describe each step of the quantum part in more details and provide a step-by-step description for the case of the H$_2$ molecule. 
    \begin{algorithm}
    \caption{Variational Quantum SCI (VQ-SCI) algorithm}\label{alg:QESCI}
    \begin{enumerate}
        \item Construct the SCI matrix (classical):
            \begin{enumerate}
                \item Select a set of \SD using any favorable SCI algorithm.
                \item Order the Slater determinants.  
                \item Construct the $M_{CI}$ matrix in the form of Eq.~\ref{eq:CI_matrix}.
            \end{enumerate}
        \item Find the groundstate of the SCI matrix (quantum):
            \begin{enumerate}
                \item Map the \SD to computational basis states. 
                \item Encode the Hamiltonian matrix to a qubit operator. 
                \item Find the $M_{CI}$ groundstate via a variational quantum circuit.
            \end{enumerate}  
    \end{enumerate}  
    \end{algorithm}

    \subsection{Finding the groundstate of the SCI matrix}
    \paragraph{Mapping the \SD to computational basis states}
    Given the ordered set of Slater determinants, we map the $k^{th}$ determinant to the $k^{th}$  computational basis state $\chi_k \!\!\rightarrow \!\!\ket{k}\!\!=\!\!\ket{b_0^kb_1^k...b_{q-1}^k}, b_j^k\in\{0,1\}$, using standard binary encoding $k\!\!=\!\!\sum_{j=0}^{q-1}2^jb_j^k$ (any other decimal-to-binary encoding will also do). This requires $q\!\!=\!\!\ceil{\log_2{(D)}}$ qubits to encode $D$ Slater determinants.

    \paragraph{Hamiltonian encoding in the computational basis}
    Using the computational basis states $\ket{k}$ the $M_{CI}$ matrix can be encoded as:
    \begin{equation}
            M_{CI} =  \sum_{j,k=0}^{D-1} M_{CI}(j,k) \ket{j}\bra{k}, 
        \end{equation} 
    with $M_{CI}(j,k) = \bra{j}M_{CI}\ket{k} = \bra{\chi_j}M_{CI}\ket{\chi_k}$, corresponding to Eq.~\ref{eq:CI_matrix}.  
    Next note that each matrix index $\ket{j}\bra{k}$ can be encoded to a qubit operator using a tensor product of $q$ single-qubit operators as follows (see e.g.~\cite{sawaya2020resource_qubit_efficient_d_level}):   
    \begin{equation}
        \ket{j}\bra{k} = \bigotimes_{n=0}^{q-1}{\ket{b_n^j}\bra{b_n^k}}
        \label{eq:entry_encoding}
    \end{equation}
    where each single qubit operator ${\ket{b_n^j}}\bra{b_n^k}$ is given by one of the following four possible single-qubit operators:
    \begin{align}
        \label{eq:entry operators}
        \ket{0}\bra{0}&=\frac{1}{2}(I+Z), \quad \ket{0}\bra{1}=\frac{1}{2}(X+iY) \\ 
        \ket{1}\bra{0}&=\frac{1}{2}(X-iY), \quad \ket{1}\bra{1}=\frac{1}{2}(I-Z) \nonumber
    \end{align}
    which can be expressed as a simple sum of two Pauli gates: 
    \begin{equation}
        \ket{b_n^j}\bra{b_n^k} = \frac{1}{2}\left(\sigma_{a} + i^p \sigma_{b}\right)_{jk}, \quad \sigma_{a} \in \{I,X\}, \;\sigma_b \in \{Z,Y\}
    \end{equation}
    Overall, the resulting qubit operator is given by:
    \begin{equation}
    \label{eq:MCI_in_qubit_form}
        M_{CI} = \sum_{k,j=0}^{D-1}M_{CI}(j,k)\bigotimes_{n=0}^{q-1}\ket{b_n^j}\bra{b_n^k} = \sum_{k,j=0}^{D-1}M_{CI}(j,k)\bigotimes_{n=0}^{q-1}\frac{1}{2}\left(\sigma_{a}^{(n)} + i^p \sigma_{b}^{(n)}\right)_{jk}
    \end{equation}
    which leads, after multiplying the $q$ parentheses to:
    \begin{equation}
        \frac{1}{2^q}\sum_{j,k=0}^{D-1}M_{CI}(j,k) \sum_{l=1}^{2^q}i^{\Tilde{p}}{\bigotimes_{n=0}^{q-1}} \sigma_l^{(n)}
    \end{equation}
    where $i^{\Tilde{p}} \in \{\pm1,\pm i\}$, and $\sigma_l^{(n)}\in\{I,X,Y,Z\}$ is a Pauli operator applied to the $n^{th}$ qubit, so that each matrix index is encoded by $2^q=D$ Pauli strings of length $q$.
    
    Note that the matrix entries, $M_{CI}(j,k)$, themselves are not encoded to the quantum circuit. Instead, they are calculated classically through Eq.~\ref{eq:CI_matrix} and either stored simply as a $D$ by $D$ matrix or calculated on demand. 
    Further, the encoding of the Hamiltonian operator to the quantum circuit is different from that employed in VQE, as it encodes matrix indices rather than fermionic operators.
     It is also different than the encoding in \cite{QEE_shee2022qubit}, in which each excitation operator is separately encoded to the circuit. This introduces constraints to the choice of configurations, which resulted with a larger number of qubits.

    Finally, given the SCI Hamiltonian matrix encoded in the computational basis, we find its groundstate by optimizing a parameterized quantum circuit in the known variational manner of Eq.~\ref{eq:parameterized_energy}.

    \subsubsection[The hydrogen molecule - a detailed example with a single qubit]{The H$_2$ molecule - a detailed example with a single qubit}
    \label{sec:H2_encoding}
    Below is a step-by-step description of the groundstate energy calculation for the $H_2$ molecule using a single qubit via VQ-SCI, as outlined in Alg.~\ref{alg:QESCI}.

\paragraph{Step I - Selecting the Slater determinants} 
The H$_2$ molecule ($N\!\!=\!\!2$ electrons) has $M\!\!=\!\!4$ spin-orbitals in the sto-3g basis set, which result with $D_{FCI}\!\!=\!\!\binom{2}{1}\binom{2}{1}\!\!=\!\!4$ possible spin-restricted configurations:
\begin{align}
    \Phi_0 = SD\{\sigma_{1s,g\downarrow},\sigma_{1s,g\uparrow}\}, \quad \Phi_2 = SD\{\sigma_{1s,u\downarrow},\sigma_{1s,g\uparrow}\}\\
    \Phi_1 = SD\{\sigma_{1s,u\downarrow},\sigma_{1s,u\uparrow}\}, \quad \Phi_3 = SD\{\sigma_{1s,g\downarrow},\sigma_{1s,u\uparrow}\} \nonumber
\end{align}
where each spin-orbital takes the \emph{\textbf{g}erade} or the \emph{\textbf{u}ngerade} spatial symmetry.  

In this simple case we can select the most dominant \SD using solely symmetry considerations (in all other cases we use a different scheme, see Sec.~\ref{secsec:SD selection}): the exact groundstate of $H_2$ is known to be of the \emph{\textbf{g}erade} symmetry, and hence cannot consist of \SD that do not obey this symmetry. In particular, single excitation determinants with \emph{\textbf{u}ngerade} symmetry must be excluded \cite{szabo2012modern}. This leaves us with only two allowed Slater determinant: $\Phi_0$ and $\Phi_1$, where $\Phi_0$ is the single-determinant HF solution and $\Phi_1$ is the double excitation state relative to it. In this case the SCI groundstate is hence identical with the FCI groundstate. 

\paragraph{Step II - Ordering the Slater determinants}
In this step each Slater determinant is mapped to a particular computational state. Here, we perform the direct mapping:
\begin{align}
\label{eq:mapping_to_computational_states_H2}
    \Phi_0 \rightarrow \ket{0}, \quad \Phi_1 \rightarrow \ket{1},  
\end{align}
with which the H$_2$ groundstate is given by  
\begin{equation}
\psi = c_0\Phi_0 + c_1\Phi_1 = c_0\ket{0} + c_1\ket{1}, \quad \lvert c_{0} \rvert ^2 + \lvert c_1 \rvert ^2 = 1.
\end{equation}

In principle, we could employ the opposite mapping, but the direct mapping of Eq.~\ref{eq:mapping_to_computational_states_H2} is better because it is less prone to measurement errors. It maps the HF solution of $\Phi_0$, which  constitutes a significant portion of the final state (this is also the case for other molecules), to the $\ket{0}$ state. An error in measuring the $\ket{0}$ state is often less probable than in measuring the $\ket{1}$ state.

\paragraph{Step III - Constructing the 2-dimensional SCI matrix} 
Constructing the 2-dimensional SCI matrix $M_{CI}$, is performed by classical methods. Here, we used psi4 \cite{smith2020psi4} to calculate it, as follows:
\begin{equation}
    \!\!M_{CI}\!=\!\! 
    \twodimmat{\bra{\Phi_0}\mathcal{H}\ket{\Phi_0}}{\bra{\Phi_0}\mathcal{H}\ket{\Phi_1}}{\bra{\Phi_1}\mathcal{H}\ket{\Phi_0}}{\bra{\Phi_1}\mathcal{H}\ket{\Phi_1}}\!\!=\!\! 
    \twodimmat{-1.8266}{0.1814}{0.1814}{-0.2596}\!\!=\!\! \twodimmat{\bra{0}\mathcal{H}\ket{0}}{\bra{0}\mathcal{H}\ket{1}}{\bra{1}\mathcal{H}\ket{0}}{\bra{1}\mathcal{H}\ket{1}}
\end{equation}
where $\mathcal{H}$ is the Hamiltonian of the H$_2$ molecule, evaluated at the equilibrium bond length of 0.745 Ang and the last equality term stems from the mapping in Eq.~\ref{eq:mapping_to_computational_states_H2} (see a similar calculation in \cite{du2010nmr} for 1.4 [au] = 0.7408 Ang bond length). 

\paragraph{Step IV - Encoding the SCI matrix to Pauli operators} 
Given the SCI matrix $M_{CI}$ in the computational basis, it is next translated to  a linear combination of Pauli operators, using the decomposition of Eq.~\ref{eq:entry operators}:
\begin{align}
    \label{eq:Pauli_decomposition_H2}
    M_{CI} &= -1.8266 \ket{0}\bra{0} + 0.1814 \ket{0}\bra{1}  + 0.1814 \ket{1}\bra{0} -0.2596 \ket{1}\bra{1} \\
           &= -1.0431\hat{I} -0.7835\hat{Z} + 0.1814\hat{X} \nonumber
\end{align}
While this step is conceptually simple, it adds a large classical computational overhead, see Sec.~\ref{sec:resource_analysis}. 

\paragraph{Step V - Finding the groundstate of the SCI matrix}
Once the H$_2$ SCI matrix is decomposed to Pauli operators, we find its lowest eigenvalue via the standard variational quantum scheme.
Since we only use a single qubit and since the SCI matrix and hence the groundstate are both real, our Anzats circuit consists only of a single qubit $R_y$ rotation:
\begin{equation}
\psi(\theta) = R_y(\theta)\ket{0} = \cos\left(\frac{\theta}{2}\right)\ket{0} + \sin\left(\frac{\theta}{2}\right)\ket{1}.
\end{equation}
For comparison, in the standard VQE with JW encoding the H$_2$ groundstate in the sto-3g basis set is represented as $\psi(\theta) = \cos(\frac{\theta}{2})\ket{1100} +  \sin(\frac{\theta}{2})\ket{0011}$ using 4 qubits, and 2 qubits are required when the parity encoding is employed \cite{o2016scalable}.

\section{Results}\label{sec:results}
\subsection{Computational setup and hyper-parameters}
We ran the proposed VQ-SCI method for the following set of molecules: H$_2$, LiH, BeH$_2$, H$_2$O, NH$_3$, and C$_2$H$_4$ in the sto-3g basis set, under spin-restriction form, as defined above, see Figs.~\ref{fig:small_molecules}-\ref{fig:h20} and Table~\ref{table:results_summary} for a concise summary. 

We computed the groundstate of each molecule using: 
(a) the full-CI (FCI) groundstate energy, calculated classically via diagonalization of the FCI matrix, using the psi4 chemistry package \cite{smith2020psi4} (dashed green line); (b) selected CI groundstate energy, calculated classically via diagonalization of the SCI matrix, using standard numerical Python diagonalization (gray diamonds);
(c) noiseless statevector classical simulations of the VQ-SCI scheme (blue circles); (d) noisy classical simulations of the VQ-SCI scheme (purple triangles) through the "$ibm\_qasm$" simulator platform using 100k shots (maximum allowed, except for H$_2$ and LiH, where we used 20k shots), where in order to have a common frame of reference we used the noise model of ``$ibmq\_santiago$" (except for $H_2$ molecule where we used ``$ibmq\_lima$"); and (e) real IBMQ hardware calculations (red squares) using 20k shots (maximum available). The specific hardware selection depended mainly on number of qubits and availability. 
In the noisy and real-hardware executions we used a standard measurement error mitigation, based on a calibration matrix \cite{endo2018practical}, as implemented by Qiskit. For calculations on real devices, the entire variational procedure was performed, rather than starting from optimal parameters found on a noisy simulator, as done sometimes in the literature. 

\begin{figure}[h]
    \centering
    \includegraphics[width=1.0\textwidth]{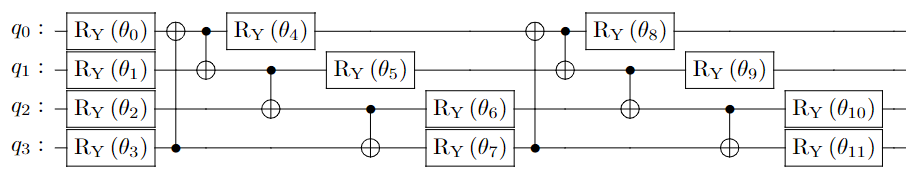}
\caption{The RealAmplitude circuit Ansatz, illustrated for the case of $q=4$ qubits and $L=2$ entangling layers. The zero'th layer consists of only $Ry$ rotations, where each further layer has $q=4$ CNOT gates and one $Ry$ rotation, per qubit.}
    \label{fig:Ansatz_circuit_example}     
\end{figure}

SCI matrices are hermitian by definition. 
Here the SCI matrices are further real (symmetric) with real eigenvectors, since they are constructed in a basis of real \SD (in the sto-3g basis set). 
This entails that the variational search can be confined to unitary transformations on the real plain. We therefore used Qiskit RealAmplitudesAnsatz (with circular entanglement), see Fig.~\ref{fig:Ansatz_circuit_example}, which is only composed from single qubit $Ry$-rotations and CNOT gates, both preserve real amplitudes of the quantum state. For each system, we set the number of circuit layers to be the minimal one with which chemical accuracy was reached using  noiseless simulations. We didn't optimize over the choice of the circuit Ansatz.  We used the COBYLA optimizer \cite{COBYLA_powell1994direct} throughout as implemented within Qiskit \cite{aleksandrowicz2019qiskit}. We didn't optimize over that choice either.

% \scalebox{0.8}{
% \Qcircuit @C=0.5em @R=0.1em @!R { \\
% 	 	\nghost{ {q}_{0} :  } & \lstick{ {q}_{0} :  } & \gate{\mathrm{R_Y}\,(\mathrm{{\ensuremath{\theta_0}}})} & \targ & \ctrl{1} & \gate{\mathrm{R_Y}\,(\mathrm{{\ensuremath{\theta_4}}})} & \qw & \qw & \targ & \ctrl{1} & \gate{\mathrm{R_Y}\,(\mathrm{{\ensuremath{\theta_8}}})} & \qw & \qw & \qw & \qw\\ 
% 	 	\nghost{ {q}_{1} :  } & \lstick{ {q}_{1} :  } & \gate{\mathrm{R_Y}\,(\mathrm{{\ensuremath{\theta_1}}})} & \qw & \targ & \ctrl{1} & \gate{\mathrm{R_Y}\,(\mathrm{{\ensuremath{\theta_5}}})} & \qw & \qw & \targ & \ctrl{1} & \gate{\mathrm{R_Y}\,(\mathrm{{\ensuremath{\theta_9}}})} & \qw & \qw & \qw\\ 
% 	 	\nghost{ {q}_{2} :  } & \lstick{ {q}_{2} :  } & \gate{\mathrm{R_Y}\,(\mathrm{{\ensuremath{\theta_2}}})} & \qw & \qw & \targ & \ctrl{1} & \gate{\mathrm{R_Y}\,(\mathrm{{\ensuremath{\theta_6}}})} & \qw & \qw & \targ & \ctrl{1} & \gate{\mathrm{R_Y}\,(\mathrm{{\ensuremath{\theta_{10}}}})} & \qw & \qw\\ 
% 	 	\nghost{ {q}_{3} :  } & \lstick{ {q}_{3} :  } & \gate{\mathrm{R_Y}\,(\mathrm{{\ensuremath{\theta_3}}})} & \ctrl{-3} & \qw & \qw & \targ & \gate{\mathrm{R_Y}\,(\mathrm{{\ensuremath{\theta_7}}})} & \ctrl{-3} & \qw & \qw & \targ & \gate{\mathrm{R_Y}\,(\mathrm{{\ensuremath{\theta_{11}}}})} & \qw & \qw\\ 
% \\ }}

The average and standard deviation of the groundstate energy were computed for noiseless simulation, noisy simulation and real hardware using the last 10 VQE iterations results, except for the $H_2O$ molecule where we ran each VQE noisy simulation 10 times and averaged the resulting energies of the last iteration, to reach more reliable results. 

In all calculations the initial state was the HF state. The optimizer convergence was usually  achieved within 50 iterations, except for the H$_2$O molecule where the real hardware computation took almost 100 iterations.

The \SD were obtained by running a restricted Hartree-Fock computation, resulting with a set of molecular orbitals, and then constructing all possible spin-restricted \SD from the set of molecular orbitals (the method, however, is not constrained to the spin-restricted form). The FCI groundstate wavefunction 
can be computed classically by common software packages. For this work we used the psi4 chemistry package to perform all of the above classical computations \cite{smith2020psi4}. Next we describe how we chose the $D_{SCI}$ most significant Slater determinants.
    
\subsection{Choosing the most significant Slater determinants} \label{secsec:SD selection}
Selecting $D$ dominant Slater determinants can be done by various classical SCI schemes, each comes with its own efficiency and accuracy \cite{zhang2020iterative,tubman2020modern,eriksen2020ground,huron1973iterative}. 
Here, however, we were not interested in the details of the selection method and therefore took a simpler route. We performed a classical FCI calculation and ranked the
configurations by their weight (absolute value of their amplitude) in the superposition groundstate. 

Next, we chose the $D_{SCI}=2^q$ most dominant (largest weight) Slater determinants by searching for the minimal $q$ which resulted with groundstate energy that reached chemical accuracy with respect to the FCI energy. In practice, we simply doubled the number of configurations, corresponding to increasing the number of qubits by 1, until chemical accuracy was achieved. This is the method we used to generate the VQ-SCI curve in Fig.~\ref{fig:qubits_per_electrons}. 
For practical calculations, when the FCI energy is not known in advance, one can increase the number of configuration until a predefined accuracy threshold is reached. 

We employed a simple decimal-to-binary mapping between \SD and the computational basis following the \SD significance order (with $\phi_0=\phi_{HF}$ being the most significant Slater determinant and $\phi_{2^q-1}$ being the least significant):
$\phi_0 \rightarrow \ket{0}^{\otimes q}, \ldots, \phi_{2^q-1} \rightarrow \ket{1}^{\otimes q}$. As already mentioned in Sec.~\ref{sec:H2_encoding},  
the advantage in mapping $\phi_0=\phi_{HF} \rightarrow \ket{0}^{\otimes q}$ is that it lessens the effect of measurement errors, as the weight of the HF state in the final groundstate is relatively high. For the same reason we also initialize the variational quantum state to $\ket{0}^{\otimes q}$.

\subsection[Small molecules: H2, LiH, and BeH2]{Small molecules: H$_2$, LiH, and BeH$_2$}
\begin{figure}[t!] 
\begin{subfigure}{0.32\textwidth}
\includegraphics[width=\linewidth]{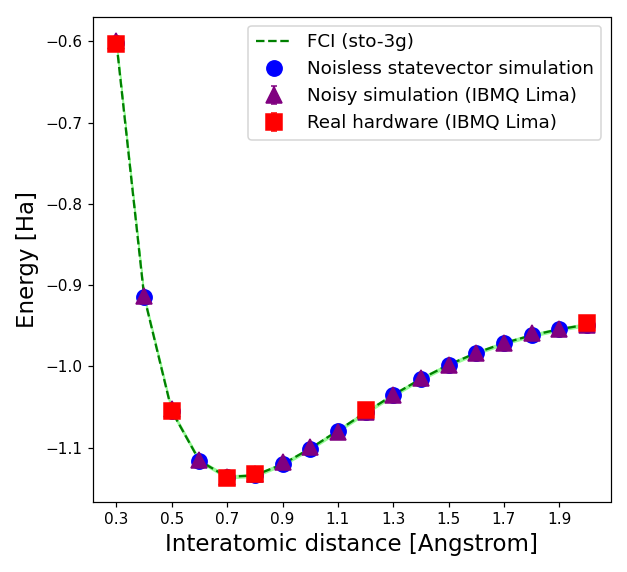}
\caption{H$_2$ potential energy}
\label{fig:H2_potential_energy}
\end{subfigure}\hspace*{\fill}
\begin{subfigure}{0.32\textwidth}
\includegraphics[width=\linewidth]{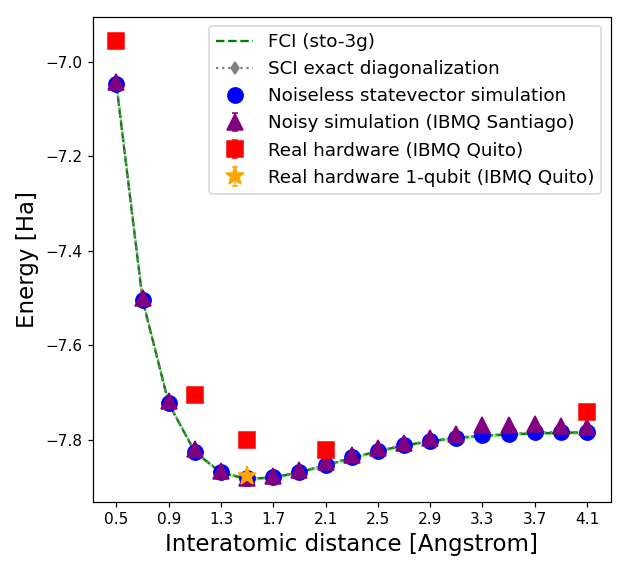}
\caption{LiH potential energy}
\label{fig:LiH_potential_energy}
\end{subfigure}
\hspace*{\fill}
\begin{subfigure}{0.32\textwidth}
\includegraphics[width=\linewidth]{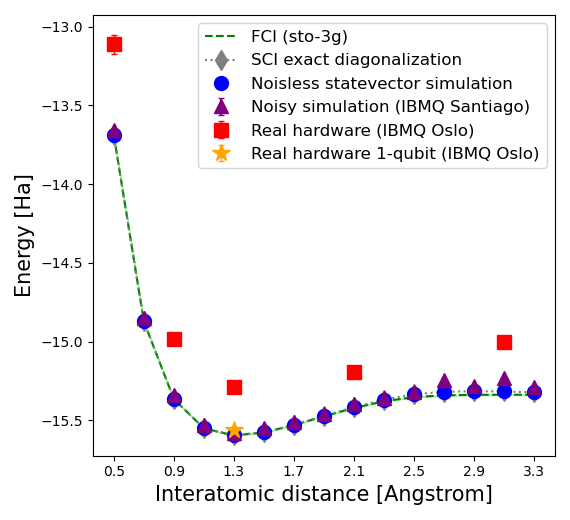}
\caption{BeH$_2$ potential energy} 
\label{fig:BeH2_potential_energy}
\end{subfigure}
\medskip
\begin{subfigure}{0.32\textwidth}
\includegraphics[width=\linewidth]{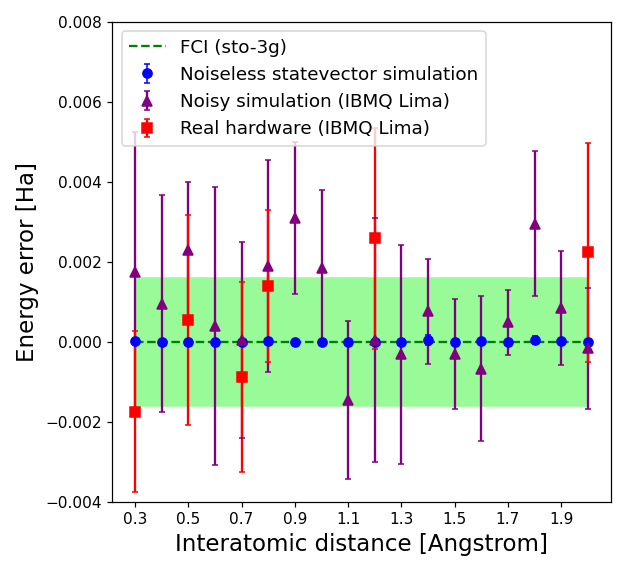}
\caption{H$_2$ energy error} \label{fig:H2_error}
\end{subfigure}
\hspace*{\fill}
\begin{subfigure}{0.32\textwidth}
\includegraphics[width=\linewidth]{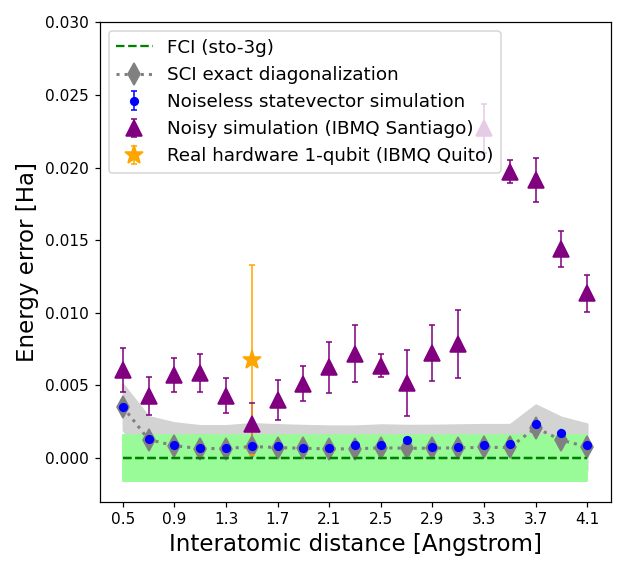}
\caption{LiH energy error} \label{fig:LiH_error}
\end{subfigure}\hspace*{\fill}
\begin{subfigure}{0.32\textwidth}
\includegraphics[width=\linewidth]{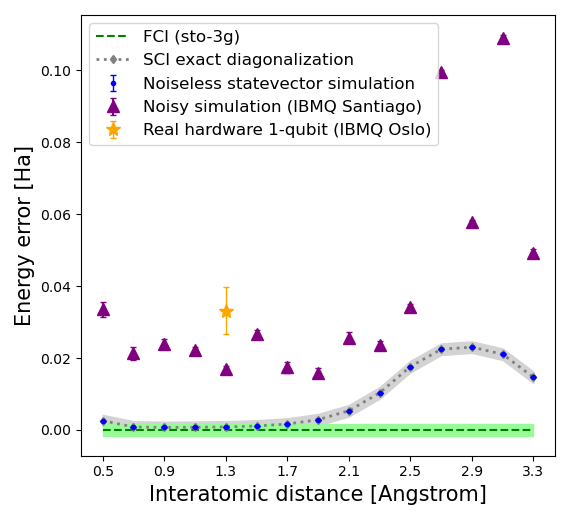}
\caption{BeH$_2$ energy error} \label{fig:BeH2_error}
\end{subfigure}
\caption{\textbf{Top row:} the VQ-SCI groundstate energy [Ha] as function of the interatomic distance [Ang], for H$_2$ (1 qubit), LiH (3 qubits), and BeH$_2$ (4 qubits). The dashed green represents the full-CI (FCI) groundstate energy; 
Gray diamonds represent the SCI groundstate energy, calculated classically via diagonalization of the SCI matrix;
Blue circles represent the SCI groundstate energy obtained by noiseless statevector classical simulations of the VQ-SCI scheme; Purple triangles indicate classical noisy simulations of the VQ-SCI scheme through the ``$ibm\_qasm$" simulator; Red squares represent real IBMQ hardware VQ-SCI calculations using 20k shots (maximum available); Finally, the yellow stars in LiH and BeH$_2$ stand for VQ-SCI calculation on real IBMQ hardware, using just a single qubit which accounts for the two most significant Slater determinants. The error bars represent the standard deviation values, observed better in the scale of the bottom row. \textbf{Bottom row:} zoom-in to the energy error [Ha] with respect to the exact FCI values, depicted using the same color scheme as above. The green area represents the chemical accuracy region (0.0016 Ha) around the exact FCI energy, whereas the gray area represents the chemical accuracy region around the exact SCI energy. For comparison, the corresponding HF groundstate energies for the H$_2$, LiH, and BeH$_2$ molecules at the equilibrium geometry are -1.1173 Ha, -7.8633 Ha, and -15.5612 Ha, respectively; the exact FCI groundstate  energies for H$_2$, LiH, and BeH$_2$ at the equilibrium geometry are -1.13618 Ha, -7.8823 Ha, and -15.59504 Ha, respectively.} 
 \label{fig:small_molecules}
\end{figure}

The top row of Fig.~\ref{fig:small_molecules} presents potential energy curves (groundstate energy as a function of interatomic distance) of the H$_2$, LiH, and BeH$_2$ molecules, as obtained with VQ-SCI, in comparison to the exact FCI results. 
To get a closer look at the method's accuracy, the bottom  row of   Fig.~\ref{fig:small_molecules} shows the deviation from the exact FCI results, as a function of interatomic distance, where the green area marks the chemical accuracy regime (0.0016 Ha) around the FCI result, the gray area marks the chemical accuracy regime around the SCI results, and the error bars indicate standard deviation. 

\paragraph[The H2 molecule]{The H$_2$ molecule - one qubit}
Calculating the groundstate of the H$_2$ molecule (2 electrons, 4 spin-orbitals) with VQ-SCI requires a single qubit, as described in Sec.~\ref{sec:H2_encoding}. This  enables the usage of the shallowest possible quantum circuit, with a single Ry rotation gate and no controllers. 
It is seen in the H$_2$ potential energy curve of  Fig.~\ref{fig:H2_potential_energy}  that the VQ-SCI energies follow closely the exact FCI curve even when executed on real device (``ibmq$\_$lima"). 
The higher resolution of the error curve in Fig.~\ref{fig:H2_error} further reveals that the noiseless statevector simulation practically coincides with the exact FCI results, and that both the noisy simulation and the real-hardware results reach chemical accuracy, up to standard deviation,  throughout. In particular, for the near equilibrium bond length of 0.7 Ang, the real-hardware grounsdtate energy is $\Delta E = \vert E-E_{FCI}\vert=0.001$ Ha from the FCI solution, well with the chemical accuracy region.  

To facilitate a clear comparison of the obtained accuracy with those reported in the literature,  Table~\ref{table:results_summary} summarizes (to the best of our knowledge) previous real-hardware groundstate calculations of the molecules studied in this work, at the equilibrium geometry, alongside with our real-hardware results.
The Table reveals that reaching chemical accuracy via VQE (or alike) on recent and current quantum devices is still not a trivial task, even for the small H$_2$ molecule in the minimal sto-3g basis set (in contrast, phase-estimation based schemes reached very high accuracy for H$_2$, much beyond chemical accuracy \cite{lanyon2010towards,du2010nmr}, but are not scalable). For example, real-hardware VQE calculations of the H$_2$ molecule reported in \cite{o2016scalable} used two qubits and reached energy that is roughly 0.02 Ha away from the exact FCI solution.
A better accuracy of 0.01 Ha was obtained in   \cite{kandala2017hardware} using a single layered circuit.
In the work of \cite{QEE_shee2022qubit} the H$_2$ molecule was computed on real quantum devices for the larger 6-31 basis set using 4 qubits, reaching an energy deviation of roughly 0.01 Ha from the corresponding FCI solution. 
In the work of \cite{jones2022chemistry} the H$_2$ groundstate was computed through the  variational quantum computed moments (QCM) method (introduced in \cite{vallury2020quantum}) using 2 qubits. The obtained groundstate energy deviated by $\Delta E = 0.001$ Ha from the FCI solution, thereby reaching chemical accuracy. 
Finally, a recent study reached chemical accuracy with standard VQE for the H$_2$ molecule via a direct pulse engineering approach, using 2 qubits \cite{meirom2022pansatz}. 
Reaching chemical accuracy was enabled in our H$_2$ calculations, despite being executed on the standard gate-based model, due to our minimal single-qubit circuit, which introduced relatively low noise.

\paragraph{The LiH molecule - 3 qubits}
The LiH molecule (4 electrons, 12 spin-orbitals) requires 8 \SD to reach chemical accuracy within the sto-3g basis set, encoded with 3 qubits. Two circuit layers were required for reaching chemical accuracy on noiseless VQ-SCI simulations, and were thus employed also in the real-hardware calculations. 
Fig.~\ref{fig:LiH_potential_energy} shows the potential energy curve of LiH as calculated by the VQ-SCI. It is seen that the noiseless and noisy calculations closely resemble the FCI curve, but that the real hardware results (``ibmq\_quito"), deviate by $\sim\!\!0.1$ Ha (with $\Delta E = $0.081 Ha at the equilibrium bond length of 1.5 Ang). 

Fig.~\ref{fig:LiH_error} presents the corresponding error energy curve. It is seen that the exact SCI results are slightly higher than the exact FCI results, but are well within the FCI chemical accuracy (except for the 3.7 Ang interatomic distance). The noiseless results are almost identical with the exact SCI diagonalization, and the noisy simulation results deviate by few mili-Hartrees from the exact SCI curve. The real-hardware results (red square) are not included in Fig.~\ref{fig:LiH_error} to maintain high resolution. It is also seen that the error (both noiseless and noisy results) increases with bond length. This is observed also in the standard VQE scheme and reflects the difficulty in accurately describing highly correlated systems, see e.g.\ \cite{fedorov2022vqe}.

For comparison with previous real-hardware groundstate calculations of LiH in the sto-3g basis set, see Table~\ref{table:results_summary}. We note that: first, in \cite{kandala2017hardware} 4 qubits were used, after freezing two spin-orbitals. The classical simulation required 6 layers, assuming an all-to-all qubit topology; Best experimental result was achieved with zero layers and reached energy of $\Delta E = \sim\!\!0.05$ Ha away from the exact FCI energy, at the 1.5 Ang equilibrium bond-length, see Fig. S9(b) in \cite{kandala2017hardware}; second, in \cite{yeter2021benchmarking} LiH was calculated with VQE using 4 qubits (the two core electrons were frozen) through symmetry preserving circuits (SPC), reaching approximately $\Delta E = 0.0025$ Ha. 
The scheme of quantum imaginary time evolution (QITE) was also employed in \cite{yeter2021benchmarking} using 2 qubits (following core electrons freezing and reduced Hamiltonian blocks), reaching groundstate energy with approximately $\Delta E = 0.005$ Ha; and finally, the recent work of \cite{QEE_shee2022qubit} on LiH (at an interatomic distance of 1.55 Ang) used 4 qubits and reached an error of approximately $\Delta E \approx 0.01$ Ha. Both \cite{yeter2021benchmarking} and  \cite{QEE_shee2022qubit} utilized the Richardson extrapolation method for noise mitigation, which we so far did not use.

We further calculated the groundstate energy at the equilibrium bond length using only the two most significant Slater determinants. This required a single qubit and a single Ry rotation gate (as in H$_2$). The real-hardware result, marked by a yellow star, is shown in Figs.~\ref{fig:LiH_potential_energy} and \ref{fig:LiH_error}. Such a calculation illustrates a trade-off between \emph{accuracy}, achieved by increasing the number of Slater determinants, and \emph{precision}, which deteriorates with increasing the number of qubits, due to the presence of more noise. In the case of LiH, SCI with two \SD deviates from the exact FCI energy by $\sim\!\!0.0055$ Ha. A similar accuracy was reproduced with real-hardware calculation. This is because of the high precision of the real-hardware single qubit calculation, as we saw before with H$_2$. Figs.~\ref{fig:LiH_potential_energy} and \ref{fig:LiH_error} indicate that in the case of LiH, the shortage of \SD is largely compensated by a significant noise reduction, reaching $\Delta E = 0.007$ Ha from the FCI result.

\paragraph[The BeH2 molecule]{The BeH$_2$ molecule - 4 qubits}
The BeH$_2$ molecule (6 electrons, 14 spin-orbitals) 
requires 16 \SD to reach chemical accuracy with respect to the FCI groundstate energy, which we encode using 4 qubits. The VQ-SCI required 3 layers to reach chemical accuracy on noiseless simulation. Using the same number of layers, real-hardware calculation at the equilibrium bond length of 1.3 Ang, reached an error of approximately 0.3 Ha, see Fig.~\ref{fig:BeH2_potential_energy}. 
Fig.~\ref{fig:BeH2_error} demonstrates that as the atoms separate the SCI with 16 \SD is no longer sufficient, as the SCI energy is no longer within the chemical accuracy region. 

In comparison with previous literature, 6 qubits were used in  \cite{kandala2017hardware} after ignoring the $2p_y, 2p_z$ orbitals per spin-channel (total of 4 spin-orbitals), freezing 2 spin-orbitals, and tapering off 2 qubits. Noiseless simulation required 16 layers assuming an all-to-all qubit topology. Best experimental result, at the equilibrium bond-length, was achieved with zero layers and reached groundstate energy which deviated by more than $\Delta E = $0.25 Ha from the exact FCI energy, see Fig.~3 in \cite{kandala2017hardware}.

As in the case of LiH, we further performed a single-qubit, real-hardware, VQ-SCI calculation of BeH$_2$ groundstate at the equilibrium bond length, accounting for only 2 Slater determinants. This calculation reached a relatively high accuracy of $\Delta\approx 0.035$ Ha, as marked by the yellow star in Figs.~\ref{fig:BeH2_potential_energy} and \ref{fig:BeH2_error}. 

\subsection{Larger molecules: water, ammonia, and ethylene}\label{secsec:larger_molecules}
\begin{figure}
    \centering
    \subfloat[\centering H$_2$O]
    {{\includegraphics[width=0.33\textwidth]{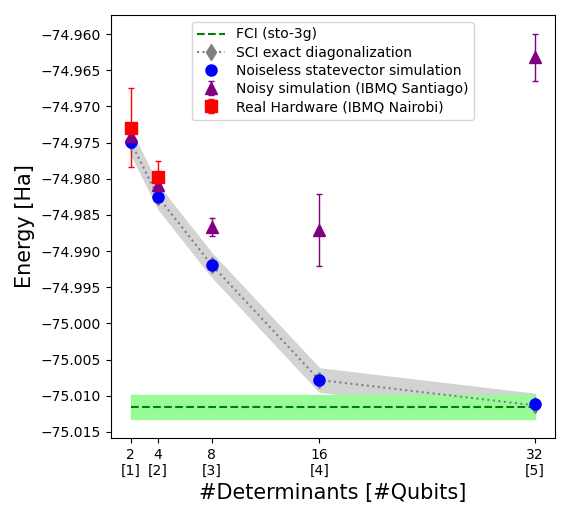} }\label{figfig:H2O}}
    \subfloat[\centering NH$_3$]{{\includegraphics[width=0.33\textwidth]{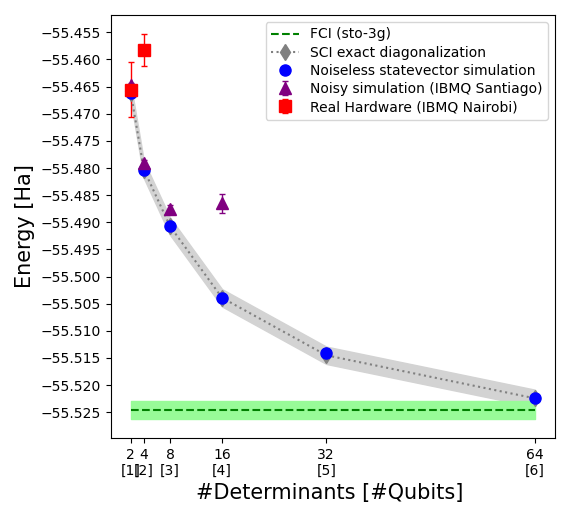} }} 
     \subfloat[\centering C$_2$H$_4$]{{\includegraphics[width=0.33\textwidth]{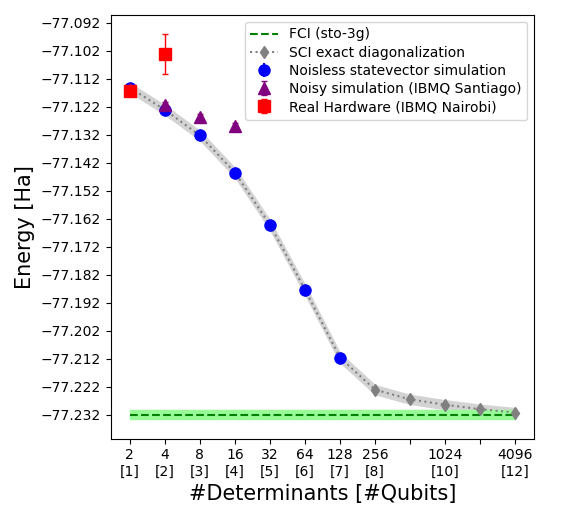} }}
     \caption{The groundstate energy [Ha] as a function of the number of significant Slater determinant configurations, corresponding to a growing number of qubits, for the H$_2$O, NH$_3$, and C$_2$H$_4$ molecules, calculated at the equilibrium geometry. The dashed green line represents the full-CI (FCI) groundstate energy, where the green area marks the chemical accuracy region (0.0016 Ha); Gray diamonds with dotted line represent the SCI groundstate energy, calculated classically via diagonalization of the SCI matrix, where the gray area marks the chemical accuracy region around the exact SCI solution; Blue circles represent the SCI groundstate energy obtained by noiseless statevector classical simulations of the VQ-SCI scheme; Purple triangles indicate classical noisy simulations of the VQ-SCI scheme through the ``$ibm\_qasm$" simulator; Red squares represent real IBMQ hardware VQ-SCI calculations using 20k shots (maximum available); The error bars represent the standard deviation values. For comparison, the corresponding HF groundstate energies for the H$_2$O, NH$_3$, and C$_2$H$_4$ molecules are -74.9624 Ha, -55.4554 Ha, and -77.07395 Ha, respectively; the exact FCI groundstate  energies for H$_2$O, NH$_3$, and C$_2$H$_4$ are -75.01156 Ha, -55.5245 Ha, and -77.2321 Ha, respectively}. 
    \label{fig:h20}
\end{figure}

Fig.~\ref{fig:h20} shows the VQ-SCI results for the H$_2$O, NH$_3$, and C$_2$H$_4$ molecules. For each molecule the convergence of the groundstate energy is shown as a function of the number of significant Slater determinants, selected as described in Sec.~\ref{secsec:SD selection}. All calculations were performed at a fixed molecular geometry, as detailed per molecule.

\paragraph[The water molecule - 5 qubits]{The H$_2$O molecule (water) - 5 qubits}
The ground-state energy of the H$_2$O molecule (10 electrons, 14 spin-orbitals) was calculated at a fixed equilibrium geometry of O-H bond-length 0.955 [Ang] and an angle of $105^\circ$, as in \cite{nam2020ground_watre_vqe_trapped_ion}. 
Applying spin-restriction to the wavefunction results with a total of 441 possible Slater determinants. Yet, as entailed by   Fig.~\ref{figfig:H2O}, 32 determinants are already sufficient for reaching the chemical accuracy (green area), thereby requiring merely 5 qubits.  
We used $\{0,1,2,3,5\}$ layers for $1-5$ qubits, respectively, except for the noisy simulation of 5 qubits for which we used only 4 layers.  
For the noisy simulation we used up to 300 iterations and 100k shots, whereas using real-hardware we used up to 100 iterations and 20k shots (due to real-hardware availability). For each number of configurations, the noisy simulation was performed 10 times to obtain more stable results, upon which we averaged and took the standard deviation.

    We calculated the groundstate energy of the water molecule in the exact same basis and geometry as done in \cite{nam2020ground_watre_vqe_trapped_ion}, which thus form a good reference point to our results. 
    The VQE calculations reported in 
 \cite{nam2020ground_watre_vqe_trapped_ion} employed the unitary coupled cluster (UCC) Ansatz, with a growing number of selected excitations. 
    Their classical simulations required 18 \SD to reach chemical accuracy, which were encoded with 11 qubits and 143 entangling gates.  In comparison, our VQ-SCI simulations recovered the FCI energy to chemical accuracy with 5 qubits and 25 CNOT gates, demonstrating the practical significance of the method.
    
    A similar trend is observed in the real-hardware calculations. The real hardware calculations reported in \cite{nam2020ground_watre_vqe_trapped_ion} were performed on an ion-trap quantum computer. They accounted for 2 (HF+1) and 4 (HF+3) Slater determinants using 2 and 3 qubits, and employed 2 and 6 XX gates, respectively, reaching precision deviations of up to 0.004 Ha from the exact SCI diagonalization curve.  
    Our real-hardware calculations for 2 and 4 \SD on IBM machines used 1 and 2 qubits, and employed zero and one CNOT gate (before transpiling), respectively. These calculation reached precision deviations of at most 0.003 Ha from the exact SCI diagonalization curve, see Fig.~\ref{figfig:H2O}.

    The VQ-SCI results are shown to be more precise than those reported in \cite{nam2020ground_watre_vqe_trapped_ion}. This is despite the fact that the superconducting qubit technology we used is more susceptible to decoherence noise than the ion-trap quantum technology and that  several hardware optimizations were carried in \cite{nam2020ground_watre_vqe_trapped_ion} in a co-design framework, with no external limitations on the number of shots.  The central reason for this relative success is the use of a shallower circuit Ansatz with fewer qubits, enabled by our encoding.

    We further note that the water molecule was also calculated previously using 5 qubits in \cite{eddins2022doubling} on real IBM hardware, to very good accuracy. This was done by using core-freezing, accounting for only 6 electrons in 10 spin-orbitals, and by employing the entanglement forging method. 

\paragraph[The ammonia molecule - 6 qubits]{The NH$_3$ molecule (ammonia) - 6 qubits}
We calculated the groundstate energy of the Ammonia molecule NH$_3$ (10 electrons, 16 spin-orbitals) at the HF (sto-3g) equilibrium geometry with N-H distance of 1.0325 Ang and H-H distance of 1.6291 Ang, as taken from \cite{johnson1999nist}. Fig.~\ref{fig:h20}(b)  shows that chemical accuracy can be reached with 64 \SD using 6 qubits, in contrast to 16  qubits that would be required in the standard VQE.  
While our noisy simulations matched the noiseless simulation quite well up to 3 qubits, the real-hardware results deviated by almost 0.02 Ha already for two qubits. 

\paragraph[The ethylene molecule]{The C$_2$H$_4$ molecule (ethylene) - 12 qubits}
\label{secsecsec:c2h4}
Ethylene (16 electron, 28 spin-orbitals) is the largest molecule we addressed. The geometry of the molecule was defined according to HF ground state energy in equilibrium geometry in the sto-3g basis, as given in \cite{johnson1999nist} (see appendix \ref{secA1:c2h4_geometry}).  Using the sto-3g minimal basis under spin-restricted form results with $\binom{14}{8}^2= 9,018,009$ \SD. Fig.~\ref{fig:h20}(c) shows the groundstate energy of ethylene as a function of the number of selected Slater determinants. It is seen that 4096 \SD are sufficient for obtaining chemical accuracy, which can be simulated with 12 qubits. 

Our noiseless simulation for 
$\{1,2,3,4,5,6,7\}$ qubits were performed with 
$\{0,1,2,3,7,11,18\}$ layers, respectively. 
We performed up to 500 iterations in all cases. Note that here we did not go beyond 7 qubits, due to the runtime consumption of our current implementation, which scales as $D^3$, but can be significantly improved, see Sec.~\ref{sec:resource_analysis}. We are not aware of previous real-hardware groundstate energy calculations of the ethylene molecule.

\begin{table}[h!]
\centering
 \begin{tabular}{||c c c c c c||} 
 \hline
 Mol. & Ref. (year) & $\#$Qubits &$\Delta$FCI (Ha) & QC & Remark \\ [0.5ex] 
 \hline\hline
 H$_2$ & \cite{lanyon2010towards} (2010) &    2 &  $10^{-6}$  & Photonic  & IPEA*\\
       & \cite{du2010nmr} (2010)         &    2 &  $10^{-13}$  & NMR  & PEA* \\
       & \cite{o2016scalable}     (2016) &    2 &  0.02  & Xmon &  PEA*\\
       & \cite{o2016scalable}     (2016) &    2 &  0.02  & Xmon & \\
       & \cite{kandala2017hardware} (2017)&    2 &  0.01  & IBM  &  \\ 
       & \cite{QEE_shee2022qubit} (2022) &    4 &  0.01  & IBM  & 6-31G basis set \\
       
       & \cite{jones2022chemistry} (2022)  &    2 &  0.001 & IBM  & QCM* \\
       & \cite{meirom2022pansatz} (2022)  &    2 &  0.001 & IBM  & Pulse engineering \\
       & This work                 &  1  & 0.001 & IBM  &\\
 \hline
 LiH   &  \cite{kandala2017hardware} (2017) & 4 & 0.05    &  IBM &  \\ [0.5ex] 
       &  \cite{yeter2021benchmarking} (2021) & 4 & 0.0025  &  IBM &  COF + EE \\
        &  \cite{yeter2021benchmarking} (2021) & 2 & 0.005   &  IBM    & QITE* + COF + EE \\ 
        & \cite{QEE_shee2022qubit} (2022)   & 4 & -0.01 & IBM  & EE   \\
       & This work                    &  3 & 0.081 & IBM &\\
       & This work*                   &  1 & 0.007 & IBM  &\\
\hline
 BeH$_2$   &  \cite{kandala2017hardware} (2017) & 6 &$0.25$ &  IBM & \\ [0.5ex]  
           & This work                  & 4 & 0.3  & IBM &\\ 
           & This work*                  & 1 & 0.035  & IBM  &\\
 \hline
 H$_2$O    &  \cite{nam2020ground_watre_vqe_trapped_ion} (2020)   & (11) 3 & 0.028    & IonQ  &   (18) 4 SDs \\ [0.5ex] 
           &  \cite{eddins2022doubling} (2022)                     & 5  & 0.007      &  IBM         & COF  \\ 
           & This work*             &    (5) 2               &  0.032    & IBM   & (32) 4 SDs \\
\hline
 NH$_3$   & This work*              &   (6) 1 &  0.06 & IBM  &\\  [0.5ex]
\hline
 C$_2$H$_4$ &  This work*             &    (12) 1 & 0.12 & IBM   & \\ [0.5ex]
\hline
\hline
 \end{tabular}
 \caption{
 A summary of groundstate calculations of the molecules studied in this paper, in equilibrium geometry, performed on real quantum hardware, as reported in the literature, and ordered chronologically. For each study we specify the number of qubits used in the experiment (numbers in parentheses indicate how many qubits are required to reach the FCI solution), the deviation from the FCI solution in Hartree,   and the quantum hardware technology that was used. The calculations were done in the sto-3g basis set, unless otherwise stated. It is evident that the \qesci consistently employs the least number of qubits, except for cases where core orbital freezing (COF) is used, which effectively accounts for a smaller number of electrons. All results from literature were obtained with VQE, unless marked with a star in the remark column. 
 The following acronyms are used: phase estimation algorithm (PEA), iterative PEA (IPEA), quantum computed moments (QCM), error extrapolation (EE), quantum imaginary-time evolution (QITE), and Slater-determinants (SDs).}
 \label{table:results_summary}
\end{table}

\section{Resource analysis and discussion}
\label{sec:resource_analysis}
As all variational quantum algorithms, the \qesci algorithm requires quantum and classical resources. On the quantum side, it requires $log D$ qubits and $O(LPSI)$ circuit executions,
where $L$ is the number of layers in the quantum circuit, $P$ is the number of distinct Pauli-strings that have to be measured, $S$ is the number of shots required to achieve a certain statistical accuracy per Pauli string measurement, and $I$ is the number of required iterations.
On the classical side, the \qesci requires computing and storing $D$ Slater determinants, the computation of $P$ Pauli strings and the optimization of $O(LlogD)$ variational parameters (one parameter per qubit times $L$ layers), per iteration. 

In terms of the required number of qubits, the \qesci method offers a significant advantage in comparison to standard VQE, as shown in Fig.~\ref{fig:qubits_per_electrons} and discussed in Sec.~\ref{sec:our_approach}. 
Yet, in terms of execution time, the \qesci scales quadratically with $D_{SCI}$, which, for large molecules, can be intractable. 
To see that, we focus on $P$, the number of Pauli strings, which accounts for most of the computational burden.  
Given $D$ \SD there are $4^q=D^2$ different Pauli strings of length $q$, which gives an upper bound for $P$. 
In the case of FCI, we have $D_{FCI}^2=\binom{M/2}{N/2}^4$ different Pauli strings (assuming spin-restriction), which, in the worst case of $M=2N$, scales exponentially with the number of spin-orbitals $M$. 
In contrast, in VQE the number of distinct Pauli strings scales  polynomially with $M$ ($\mathcal{O}(M^4)$). In the case of SCI, the relation between $D_{SCI}$ and $M$ and $N$ depends on the particular selection method we use. Empirically  we observe that in our case $D_{SCI}\approx\sqrt{D_{FCI}}=\binom{M/2}{N/2}$,  
which reduces the number of Pauli strings quadratically. 

\begin{figure}[h!]
\centering
\includegraphics[scale=0.45]{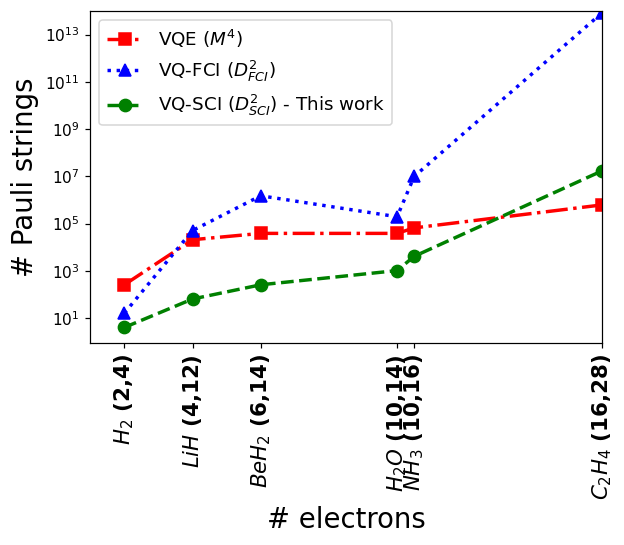}
\caption{The number of Pauli strings required to perform groundstate energy calculations of the molecules studied in this paper, ordered by their number of electrons (the NH$_3$ molecule is slightly shifted for clarity), in the sto-3g basis set, per method.}
\label{fig:PauliStringsPerSys}
\end{figure}

Fig.~\ref{fig:PauliStringsPerSys} shows the improvement of \qesci over the \vqfci in terms of the required number of Pauli strings. Moreover, it demonstrates that for a large set of molecules the \qesci requires the measurement of fewer Pauli strings than VQE. Yet, it still suffers from an intractable scaling. This can be partially remedied (so far we only considered an upper bound for $P$) by taking the hermiticity of the Hamiltonian into account and by employing methods such as the tensor product basis (TPB) method \cite{mcclean2016theory,kandala2017hardware, verteletskyi2020measurement,gokhale2019minimizing}.

It is also instructive to consider the required number of Pauli strings for an increasing basis set, i.e.\ the case where the number of electrons $(N)$ is kept fixed, while the number of spin-orbitals $(M)$ increases. Using the estimated relation $D_{SCI}\approx\binom{M/2}{N/2}$ and relying on $\lim_{K\rightarrow \infty}\binom{K}{L} \approx \frac{K^L}{L!}$, we get that
$D_{SCI}\approx \frac{(M/2)^{N/2}}{(N/2)!}$ 
and $P_{SCI} \approx \frac{(M/2)^{N}}{((N/2)!)^2}$.
For small molecules or for calculations with small active space (small $N$), calculated in large basis sets (large $M$) this can be advantageous in comparison to VQE, as demonstrated in Appendix \ref{secA2:large basis sets}. Note further that in this case the number of qubits required by \qesci is exponentially smaller than that required by VQE ($M$), since $log(D_{SCI})\approx \frac{N}{2}log\frac{M}{2}$, as also observed in \cite{QEE_shee2022qubit}.

The \qesci offers great flexibility. It can directly choose any subset of Slater determinants, thereby reducing the number of required qubits, while still using relatively shallow circuits.  
The usage of shallower circuits effects both the induced noise as well as the number of overall circuit executions. This is because each variational parameter in the circuit Ansatz requires further circuit executions for the sake of the optimization process. 
Our resource analysis neglected the effect of the actual number of variational parameters in the quantum circuit. This number is relevant for the overall execution time and can be largely reduced by the reduction in the number of qubits, thereby reducing the actual runtime gap between VQE and VQ-SCI.

\subsection{Mitigating the classical memory bottleneck}
One of the most severe obstacles in classical SCI computation is the memory bottleneck \cite{zhang2020iterative,tubman2020modern}. Most often an explicit storage of the SCI matrix is avoided by calculating it on the fly. Similarly, keeping an explicit list of the \SD can be avoided, especially when the selection is done \emph{a-priory}, by a fixed rule, such as in CISD, see \cite{garniron2018development} and references therein. The central memory (and time) bottleneck originates in finding the groundstate of the $D$-dimensional $M_{SCI}$ matrix, where $D=D_{SCI}$. Classical SCI schemes employ iterative schemes such as the Davidson algorithm which finds the groundstate of a given $D$-dimensional matrix in $\mathcal{O}(D^2)$ time and $\mathcal{O}(D)$ memory \cite{DAVIDSON197587}. As $D$ can reach the order of $10^{12}$ determinants, see e.g. \cite{vogiatzis2017pushing},  
the classical computation is often parallelized between many cores, which has a high communication cost.   

The \qesci offers a bypass to this classical memory bottleneck. 
This can be done by computing non-zero matrix elements sequentially and generating the Pauli strings one by one, per matrix entry, on the fly. This will exponentially 
 reduce the classical storage requirement to $log D$ for holding only the $log D$ 2-qubit gates that encode the relevant matrix index-operator, see Eq.~\ref{eq:entry_encoding}. Importantly, this exponential improvement in memory consumption can only be achieved when running the \qesci on real devices. 
Simulating it classically will not benefit from calculating the Pauli strings on the fly. It would still require $2^q=D$ time and memory just to simulate the $q$-qubits circuit.

Reducing the classical memory comes at a computational time cost, as some Pauli strings will be measured multiple times. However,
this approach is particularly suitable for the SCI matrix, which is very sparse. According to the Slater–Condon rules the matrix entries  $M_{SCI}(j,k)$ (see Eq.~\ref{eq:CI_matrix}) vanish whenever the \SD $\Phi_j$ and $\Phi_k$ differ by more than two spin-orbitals \cite{szabo2012modern}. This implies that each row has merely  
$\binom{N}{1}\binom{M-N}{1} + \binom{N}{2}\binom{M-N}{2} < N^2(M-N)^2 << D$
non-zero elements (in the spin un-restrictive case),  
where the exact sparsity will be dictated by the specific determinants selection (which may be initially  done so as to increase sparsity). 
In conclusion, for a sparse $D$-dimensional matrix with  $\mathcal{O}(D)$ non-zero entries (sparsity) and $D$ Pauli strings per entry, the VQ-SCI method requires $\mathcal{O}(D^2)$ computational time (similar to the Davidson algorithm), but with $log(D)$ memory (compared to $\mathcal{O}(D)$ in the case of Davidson algorithm).

\section{Conclusion and outlook}\label{sec:conclusion}
We developed a framework for performing selected configuration interaction calculation through a variational quantum algorithm approach, denoted VQ-SCI. With this method we were able to simulate ground state calculations to FCI chemical accuracy for a set of small to medium size molecules, with the smallest to date number of qubits. 
The method uses a hardware efficient approach, while preserving the number of particles, and was shown to reach relatively high accuracy on IBM quantum devices, even without using advanced error mitigation schemes. Specifically, we reached chemical accuracy for the $H_2$ molecule through a single qubit calculation, and performed real-hardware groundstate calculations of ethylene for the first time, using a small selected subset of significant Slater determinants.

 The number of distinct Pauli strings that need to be measured in \qesci scales worse than in VQE but can potentially outperform VQE for small-to-medium size molecules. 
 Moreover, even for SCI of larger molecules, the overall number of circuit executions may eventually be larger in VQE due to the larger number of variational parameters it requires, in practice. 
 For molecules for which the number of circuit executions is larger than in VQE, it may nevertheless be more feasible, simply due to its better noise resistance. 
The \qesci may therefore be particularly valuable for the NISQ era, where qubits are scarce and noise is high. 
To reduce its time complexity,  further research is required to minimize the number of needed measurements. To that end, it is worthwhile to: (a) explore the usage of Pauli grouping methods, such as TPB; (b) clarify the impact of other binary encodings, such as the Gray code, on the required resources; (c) find matrices which can be decomposed to small number of Pauli strings. For practical usage it is also beneficial to use more advanced error mitigation schemes.

In comparison to classical SCI methods, the \qesci scales similarly in time, but offers an exponential reduction in memory consumption, which is a severe bottleneck in such classical computations. In that sense the \qesci can potentially offer a practical advantage over classical eigensolvers, such as the Davidson algorithm.

Finally, the \qesci 
provides a general quantum route for finding the groundstate of Hermitian matrices and can be used outside the chemical context, exhibiting the same advantage in memory scaling.

\bmhead{Acknowledgments}
We acknowledge the use of IBM Quantum services for this work. The views expressed are those of the authors, and do not reflect the official policy or position of IBM or the IBM Quantum team.

\bmhead{Code availability}
The underlying code for this study is available in GitHub and can be accessed via this link \url{https://github.com/danieli4444/SCI_VQE}.

\begin{appendices}

\section[Ethylene geometry]{C$_2$H$_4$ geometry}\label{secA1:c2h4_geometry}
For the C$_2$H$_4$ calculation we used the following HF equilibrium geometry, taken from  NIST CCCBDB database \cite{johnson1999nist}:
\begin{figure}[h!]
\centering
\includegraphics[scale=0.7]{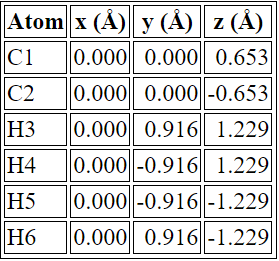}
\caption{C$_2$H$_4$ equilibrium geometry corresponding to the HF solution.}
\label{fig:C2H4_NIST}
\end{figure}

\section{Large basis sets}\label{secA2:large basis sets}
\begin{figure}[h!]
    \centering
    \subfloat[\centering H$_2$ ($\#$electrons = 2)]{{\includegraphics[width=0.33\textwidth]{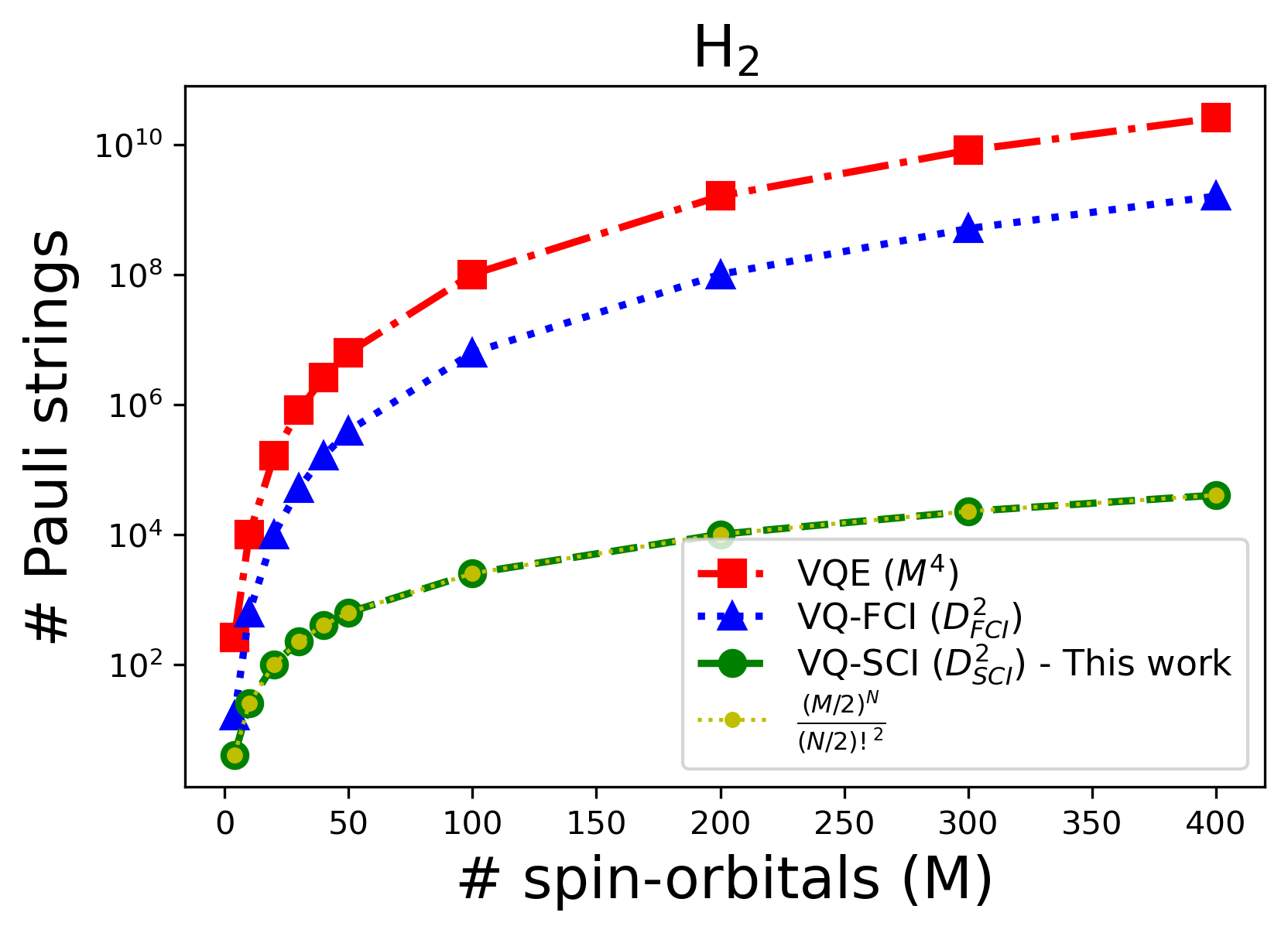} }}
    \subfloat[\centering LiH ($\#$electrons = 4)]{{\includegraphics[width=0.33\textwidth]{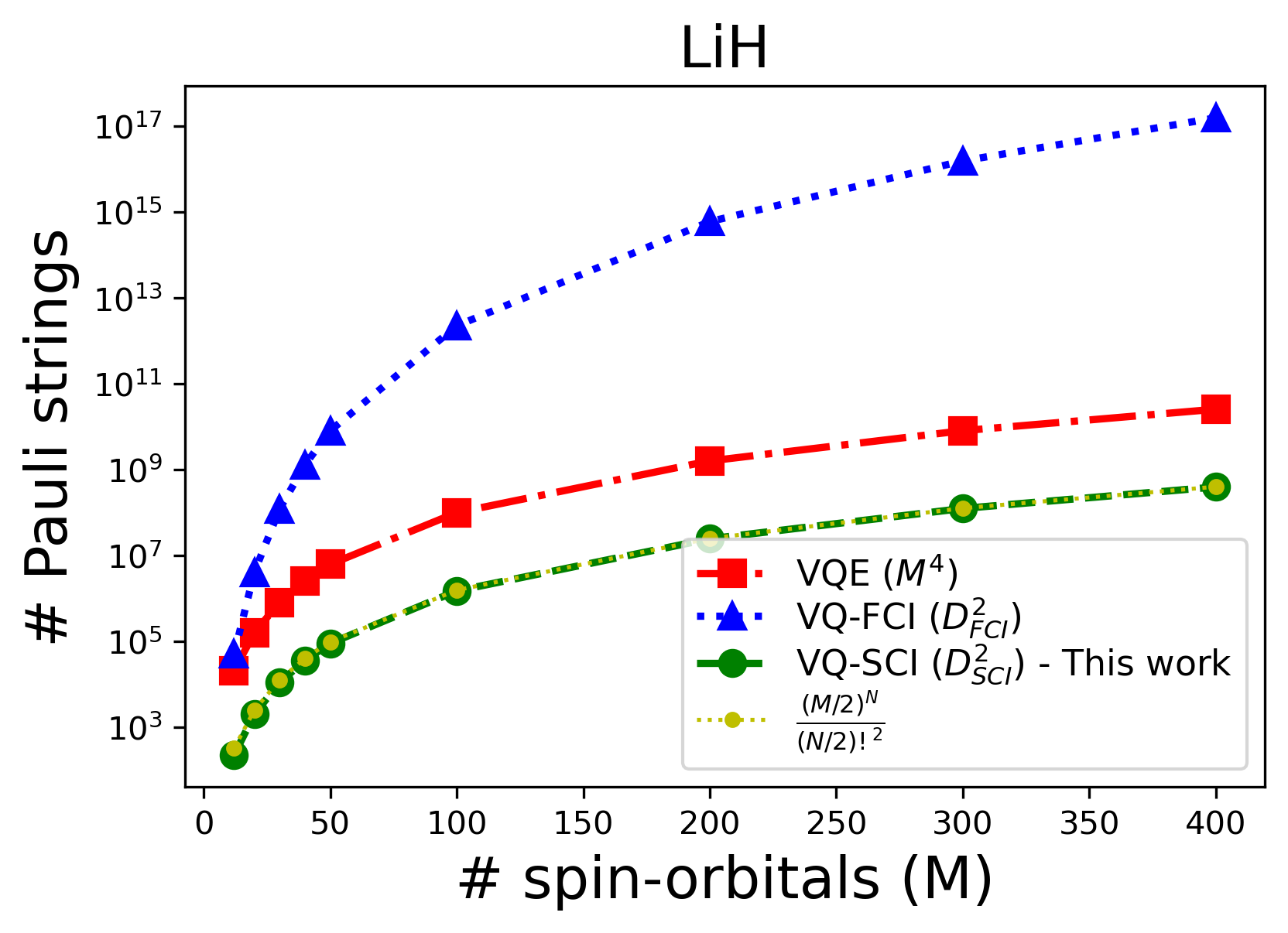} }} 
     \subfloat[\centering BeH$_2$ ($\#$electrons = 6)]{{\includegraphics[width=0.33\textwidth]{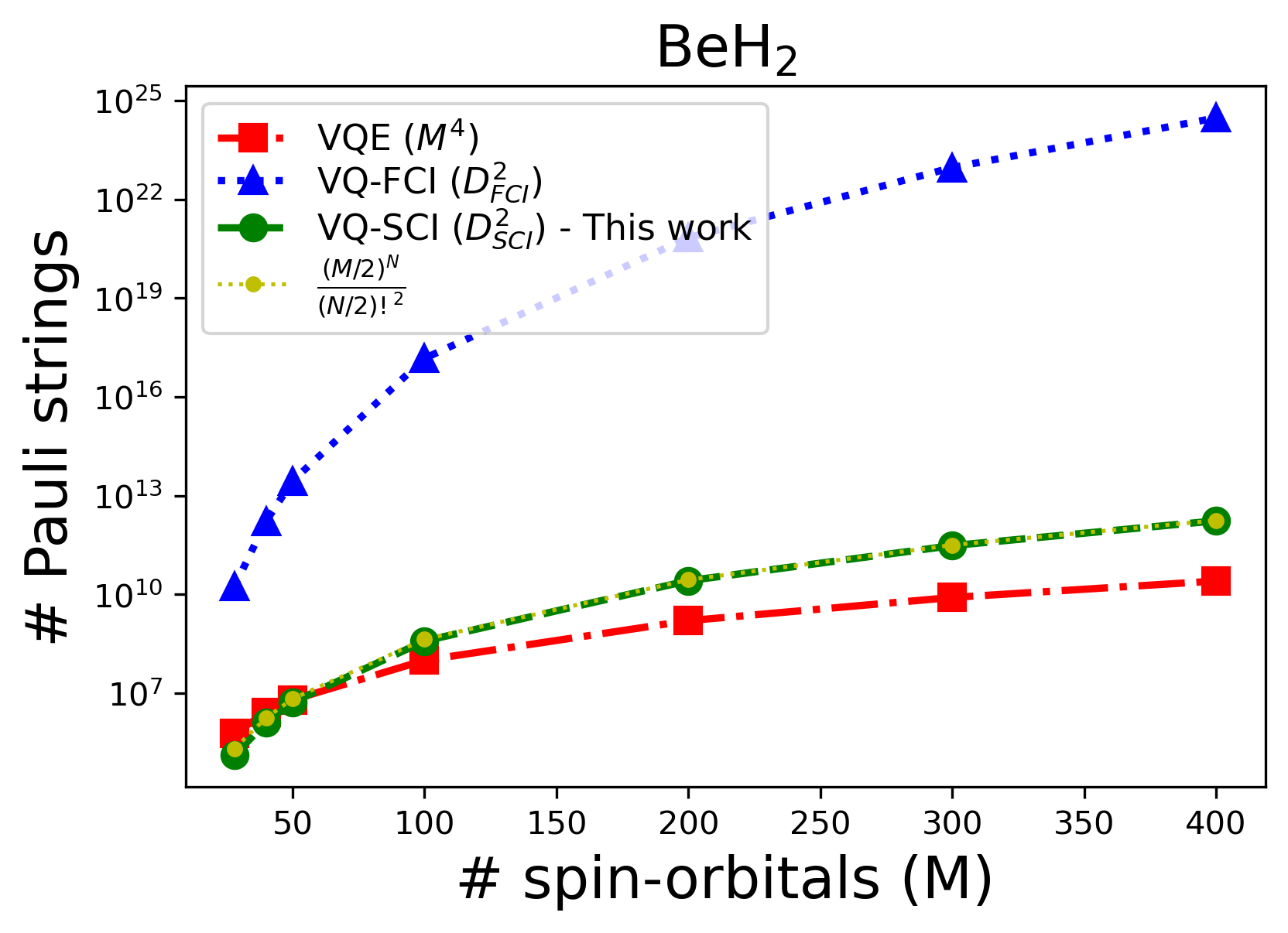} }}
    \caption{The number of Pauli strings required per method, as a function of the number of spin-orbitals $(M)$, for the H$_2$, LiH, and BeH$_2$ molecules.}
\label{fig:Pauli_strings_fixed_molecule_large_basis}
\end{figure}

Fig.~\ref{fig:Pauli_strings_fixed_molecule_large_basis} shows the scaling of the required number of distinct Pauli strings when increasing the size of the basis set, namely when the number of electrons $(N)$ is kept fixed, while the number of spin-orbitals $(M)$ increases. It is seen that for small molecules in large basis sets, the VQ-SCI provides an advantage over VQE also in that respect. 
The yellow dotted line represent the estimation of $P_{SCI}=\frac{(M/2)^N}{(N/2)!^2}$ which is derived in the main text,  based on the approximated relation $D_{SCI}=\sqrt{D_{FCI}}$. A very good fit is seen with the green curve, which estimates the number of Pauli strings with $P_{SCI}=D_{SCI}^2$, based on actual number of Slater determinants we used ($D_{SCI}$).

\end{appendices}

\bibliography{sn-bibliography}% common bib file
\end{document}